\documentclass{cernrep}

\usepackage{bm}
\usepackage{amsmath}
\usepackage{amssymb}
\usepackage{graphicx}
\usepackage{url}

\usepackage[displaymath]{lineno}

\renewcommand{\i}{\mathrm{i}}

\newcommand{\fe}{\mathbf{\tilde{E}}}
\newcommand{\fb}{\mathbf{\tilde{B}}}
\newcommand{\fj}{\mathbf{\tilde{J}}}

\newcommand{\fk}{\mathbf{k}}
\newcommand{\fkhat}{\mathbf{\hat{k}}}

\renewcommand{\vec}[1]{\boldsymbol{#1}}
\newcommand{\vgal}{\vec{v}_{gal}}
\newcommand{\nab}{\vec{\nabla'}}
\newcommand{\Dt}[1]{ \frac{\partial #1}{\partial t}}
\newcommand{\mc}[1]{\hat{\mathcal{#1}}}

\usepackage{fancyhdr}
\fancyhfoffset{4 mm}
\fancypagestyle{ARTTITLE}{%
\fancyhf{} 
\lhead{\small{Proceedings of the 2019 CERN--Accelerator--School course on\\
      \it{ High Gradient Wakefield Accelerators}, Sesimbra, (Portugal)}} 
\lfoot{Available online at \url{https://cas.web.cern.ch/previous-schools}}
\rfoot{\thepage\hspace*{3mm}}
 
   }

\usepackage[bookmarks, colorlinks=true, linktoc=page, linkcolor=black, citecolor=black, urlcolor=blue]{hyperref}

\begin{document}
\title{Simulation of plasma accelerators with the Particle-In-Cell method}
\author{J.-L. Vay\thanks
                 {jlvay@lbl.gov}}
\institute{Lawrence Berkeley National Laboratory, Berkeley, USA}

\begin{abstract}

We present the standard electromagnetic Particle-in-Cell method, 
starting from the discrete approximation of derivatives 
on a uniform grid. The application to second-order, centered, finite-difference 
discretization of the equations of motion and of Maxwell's equations is 
then described in one dimension, followed by two and three dimensions.
Various algorithms are presented, for which we discuss the stability and 
accuracy, introducing and elucidating concepts like ``numerical 
stochastic heating'', ``CFL limit'' and ``numerical dispersion''.
The coupling of the particles and field quantities via interpolation 
at various orders is detailed, together with its implication on energy 
and momentum conserving. 
Special topics of relevance to the modeling of plasma accelerators  
are discussed, such as moving window, optimal Lorentz boosted frame, the 
numerical Cherenkov instability and its mitigation.
Examples of simulations of laser-driven and particle beam-driven 
accelerators are given, including with mesh refinement. 
We conclude with a discussion on high-performance computing and a brief outlook.

\end{abstract}


\maketitle
\thispagestyle{ARTTITLE}

\section{Introduction}

Computer simulations have had a profound impact on the design and understanding of past and present plasma acceleration experiments \cite{TsungPoP06,GeddesJP08,GeddesSciDAC09,HuangSciDAC09}, with  
accurate modeling of wake formation, electron self-trapping and acceleration requiring fully kinetic methods (usually Particle-In-Cell) using large computational resources (due to the wide range of space and time scales involved). Numerical modeling complements and guides the design and analysis of advanced accelerators, and can reduce development costs significantly. Despite the major recent experimental successes\cite{LeemansPRL2014,Blumenfeld2007,BulanovSV2014,Steinke2016,Gonsalves2019}, the various advanced acceleration concepts need significant progress to fulfill their potential.  To this end, large-scale simulations will continue to be a key component toward reaching a detailed understanding of the complex interrelated physics phenomena at play. 

For such simulations,
the most popular algorithm is the Particle-In-Cell (or PIC) technique,
which represents electromagnetic fields on a grid and particles by
a sample of macroparticles. 
These simulations are extremely computationally intensive, due to the need to resolve the evolution of a driver (laser or particle beam) and an accelerated beam into a structure that is orders of magnitude longer and wider than the accelerated beam.
Hence, various techniques or reduced models have been developed to allow multidimensional simulations at manageable computational costs: quasistatic approximation \cite{SpranglePRL90,AntonsenPRL1992,KrallPRE1993,Morapop1997,Quickpic}, 
ponderomotive guiding center (PGC) models \cite{AntonsenPRL1992,KrallPRE1993,Quickpic,BenedettiAAC2010,CowanJCP11}, simulation in an optimal Lorentz boosted frame \cite{VayPRL07,Bruhwileraac08,Vayscidac09,Vaypac09,Martinspac09,VayAAC2010,MartinsNaturePhysics10,MartinsPoP10, MartinsCPC10, VayJCP2011,VayPOPL2011,VayPOP2011,Yu2016}, 
expanding the fields into a truncated series of azimuthal modes
\cite{godfrey1985iprop,LifschitzJCP2009,DavidsonJCP2015,Lehe2016,Andriyash2016}, fluid approximation \cite{KrallPRE1993,ShadwickPoP09,BenedettiAAC2010} and scaled parameters \cite{CormierAAC08,GeddesPAC09}. 

It is beyond the scope of the present report to review all or even most of the methods that have been developed. 
We will focus on the detailed presentation of the standard electromagnetic Particle-In-Cell method, as it is the most 
complete (as based on the first-principle Maxwell and particle motion equations) and versatile (as it can be applied 
to plasma and beam simulations way beyond plasma acceleration). For completeness, however, the quasistatic and 
ponderomotive guiding center models are given in Appendix \ref{Sec:QS} and \ref{Sec:PGC}.

\section{The electromagnetic Particle-In-Cell method}

\begin{figure}[ht]
\begin{center}
\includegraphics[trim={1.cm 6cm 1cm 5cm},clip,scale=0.6]{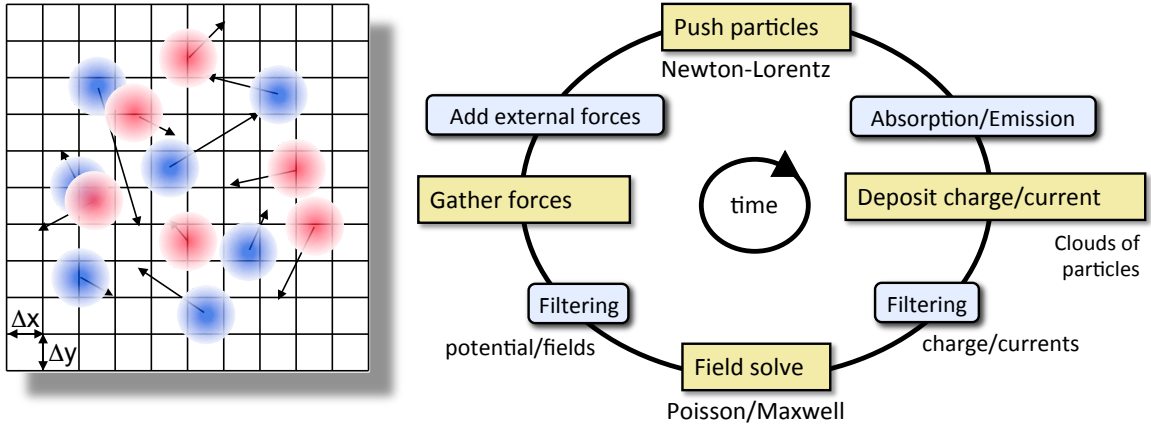}
\caption{\label{fig:PIC} The Particle-In-Cell (PIC) method follows the evolution of a collection of charged macro-particles (positively charged in blue on the left plot, negatively charged in red) that evolve self-consistently with their electromagnetic (or electrostatic) fields. The core PIC algorithm involves four operations at each time step: 1) evolve the velocity and position of the particles using the Newton-Lorentz equations, 2) deposit the charge and/or current densities through interpolation from the particles distributions onto the grid, 3) evolve Maxwell's wave equations (for electromagnetic) or solve Poisson's equation (for electrostatic) on the grid, 4) interpolate the fields from the grid onto the particles for the next particle push. Additional ``add-ons'' operations are inserted between these core operations to account for additional physics (e.g. absorption/emission of particles, addition of external forces to account for accelerator focusing or accelerating component) or numerical effects (e.g. smoothing/filtering of the charge/current densities and/or fields on the grid).}
\end{center}
\end{figure}

In the electromagnetic Particle-In-Cell method \cite{BirdsallLangdon},
the electromagnetic fields are solved on a grid, usually using Maxwell's
equations

\begin{subequations}
\begin{eqnarray}
\frac{\mathbf{\partial B}}{\partial t} & = & -\nabla\times\mathbf{E}\label{Eq:Faraday-1}\\
\frac{\mathbf{\partial E}}{\partial t} & = & \nabla\times\mathbf{B}-\mathbf{J}\label{Eq:Ampere-1}\\
\nabla\cdot\mathbf{E} & = & \rho\label{Eq:Gauss-1}\\
\nabla\cdot\mathbf{B} & = & 0\label{Eq:divb-1}
\end{eqnarray}
\end{subequations}
given here in natural units ($\epsilon_0=\mu_0=c=1$), where $t$ is time, $\mathbf{E}$ and
$\mathbf{B}$ are the electric and magnetic field components, and
$\rho$ and $\mathbf{J}$ are the charge and current densities. The
charged particles are advanced in time using the Newton-Lorentz equations
of motion 
\begin{subequations}
\begin{align}
\frac{d\mathbf{x}}{dt}= & \mathbf{v},\label{Eq:Lorentz_x-1}\\
\frac{d\left(\gamma\mathbf{v}\right)}{dt}= & \frac{q}{m}\left(\mathbf{E}+\mathbf{v}\times\mathbf{B}\right),\label{Eq:Lorentz_v-1}
\end{align}
\end{subequations}
where $m$, $q$, $\mathbf{x}$, $\mathbf{v}$ and $\gamma=1/\sqrt{1-v^{2}}$
 are respectively the mass, charge, position, velocity and relativistic
factor of the particle given in natural units ($c=1$). The charge and current densities are interpolated
on the grid from the particles' positions and velocities, while the
electric and magnetic field components are interpolated from the grid
to the particles' positions for the velocity update.

If Eq. (\ref{Eq:Gauss-1}) and Eq. (\ref{Eq:divb-1}) are verified at a given time, they are also 
verified at any later time provided that Eq. (\ref{Eq:Faraday-1}) and Eq. (\ref{Eq:Ampere-1}) 
are verified together with the continuity equation $\frac{\partial \rho}{\partial t}+\nabla \mathbf{J}=0$.
Hence Eq. (\ref{Eq:Gauss-1}) and Eq. (\ref{Eq:divb-1}) are often not solved explicitly in many codes.

\subsection{Discretization of differential operations}
A Taylor expansion of a function $f$ defined on a uniform grid of cell size $\Delta x$ writes
\begin{subequations}
\begin{eqnarray}
f_{i+1} &= & f_i + \left.\frac{\partial f}{\partial x}\right\rvert_i \Delta x 
                       + \frac{1}{2!} \left.\frac{\partial^2 f}{\partial x^2}\right\rvert_i \Delta x^2
                       + \frac{1}{3!} \left.\frac{\partial^3 f}{\partial x^3}\right\rvert_i \Delta x^3
                       + ... \;\;\; \text{    (forward direction)}, \\
f_{i+1} &= & f_i - \left.\frac{\partial f}{\partial x}\right\rvert_i \Delta x 
                       + \frac{1}{2!} \left.\frac{\partial^2 f}{\partial x^2}\right\rvert_i \Delta x^2
                       - \frac{1}{3!} \left.\frac{\partial^3 f}{\partial x^3}\right\rvert_i \Delta x^3
                       + ... \;\;\; \text{    (backward direction)}, 
\end{eqnarray}
\end{subequations}
leading to the following first order approximations of the first derivatives
\begin{subequations}
\begin{eqnarray}
\left.\frac{\partial f}{\partial x}\right\rvert_i & = & \frac{f_{i+1} - f_i }{\Delta x} + \mathcal{O}(\Delta x)\;\;\; \text{    (forward direction)},\\
\left.\frac{\partial f}{\partial x}\right\rvert_i & = & \frac{f_{i} - f_{i-1} }{\Delta x} + \mathcal{O}(\Delta x)\;\;\; \text{    (backward direction)}
\end{eqnarray}
\end{subequations}
Substracting these two first-order approximations leads to the second-order centered finite-difference approximation
\begin{equation}
\left.\frac{\partial f}{\partial x}\right\rvert_i = \frac{f_{i+1} - f_{i-1} }{2\Delta x} + \mathcal{O}(\Delta x^2)
\end{equation}

\subsection{Particle push}

\subsubsection{Simple particle integrator}

Neglecting relativity and magnetic fields, the motion equations 
\begin{subequations}
\begin{align}
\frac{d\mathbf{v}}{dt}= & \frac{q\mathbf{E}}{m} = \frac{\mathbf{F}}{m} ,\label{Eq:Motion_v-1}  \\
\frac{d\mathbf{x}}{dt}= & \mathbf{v},\label{Eq:motion_x-1}
\end{align}
\end{subequations}
can be discretized - using the second-order centered finite-difference approximation - as 
\begin{subequations}
\begin{eqnarray}
\mathbf{v}^{n+1/2} = & \mathbf{v}^{n-1/2} + \Delta t \; \mathbf{F}^{n}/m \;\;\; [+\mathcal{O}(\Delta t^2)], \\
\mathbf{x}^{n+1} = & \mathbf{x}^n + \Delta t \; \mathbf{v}^{n+1/2} \;\;\; [+\mathcal{O}(\Delta t^2)].
\end{eqnarray}
\end{subequations}
\begin{figure}[ht]
\begin{center}
\includegraphics[trim={11.5cm 3cm 1cm 10.5cm},clip,scale=0.7]{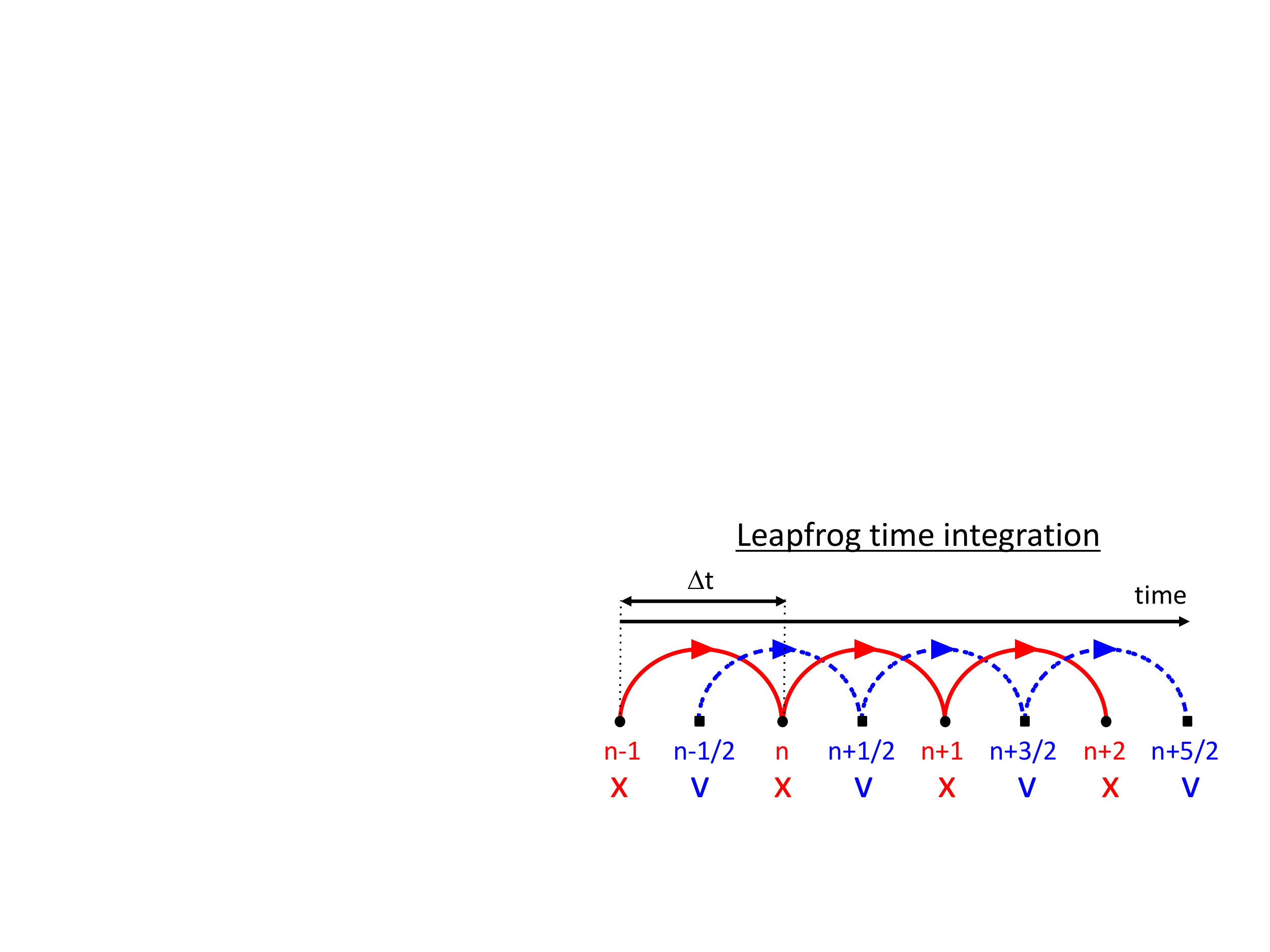}
\caption{\label{fig:leapfrog} Time integration using a second-order 
 finite-difference "leapfrog" integrator.}
\end{center}
\end{figure}
In this case, the quantities were staggered by half a time step, such that all finite-differences are centered and second-ordered. 
Such a scheme, which is represented graphically in Fig. \ref{fig:leapfrog}, is called a ``leapfrog'' time integrator.

\subsubsubsection{Stability}

Let us consider the simple case of a linear force of the form $\mathbf{F} = -m \Omega^2 \mathbf{x}$, 
i.e. of a linear oscillator
\begin{equation}
  \frac{d^2\mathbf{x}}{dt^2} = \frac{d\mathbf{v}}{dt}= -\Omega^2 \mathbf{x}.
\end{equation}
The leapfrog discretization of this equation leads to 
\begin{equation}
  \mathbf{x}^{n+1} - 2\mathbf{x}^n + \mathbf{x}^{n-1} = -\Delta t^2 \Omega^2 \mathbf{x}^n,
\end{equation}
on which we perform a Von Neumann stability anaIysis. Inserting the ansatz $\mathbf{x}=\mathbf{x}_0\exp(i\omega t)$ into the last equation gives
\begin{equation}
  \sin(\frac{\omega\Delta t}{2}) = \pm \frac{\Omega\Delta t}{2}.
\end{equation}
\begin{figure}[ht]
\begin{center}
\includegraphics[trim={5.cm 5.5cm 6cm 6.5cm},clip,scale=0.7]{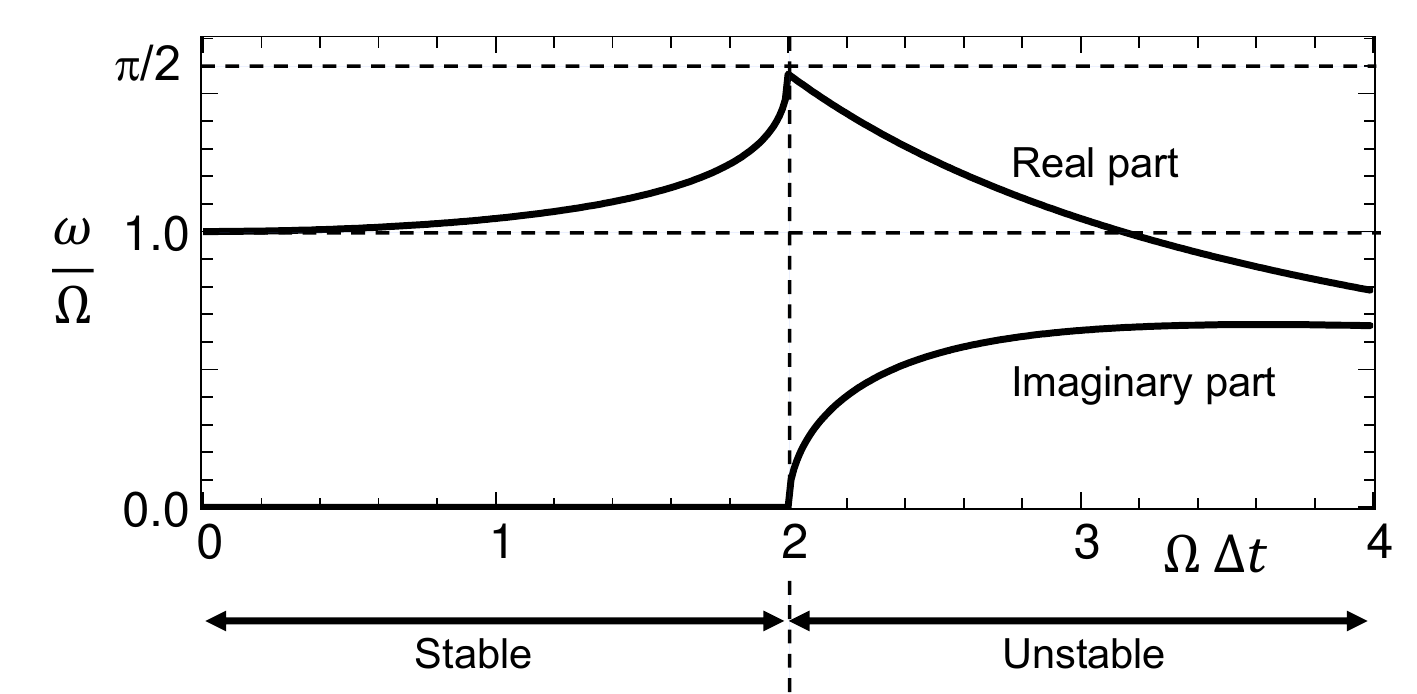}
\caption{\label{fig:leapfrogstability} Stability of a second-order 
 finite-difference "leapfrog" integrator.}
\end{center}
\end{figure}
The solution to the last expression is plotted in Fig. \ref{fig:leapfrogstability}. For $\Omega\Delta t<2$, the scheme is stable. For the entire stable region, the frequency of the oscillations is higher than the real solution. The magnitude of the error on the frequency vanishes at $\Delta t=0$ and grows with $\Delta t$. For $\Omega\Delta t>2$, the scheme is unstable. The metastable solution at $\Omega\Delta t=2$ is called the Courant-Friedrichs-Lewy, or CFL, condition, which is one of the very common limits (but not the only one) on the time step that can be used in time-dependent computer simulations.

\begin{figure}[ht]
\begin{center}
\includegraphics[trim={3.5cm 4.5cm 5cm 4.5cm},clip,scale=0.6]{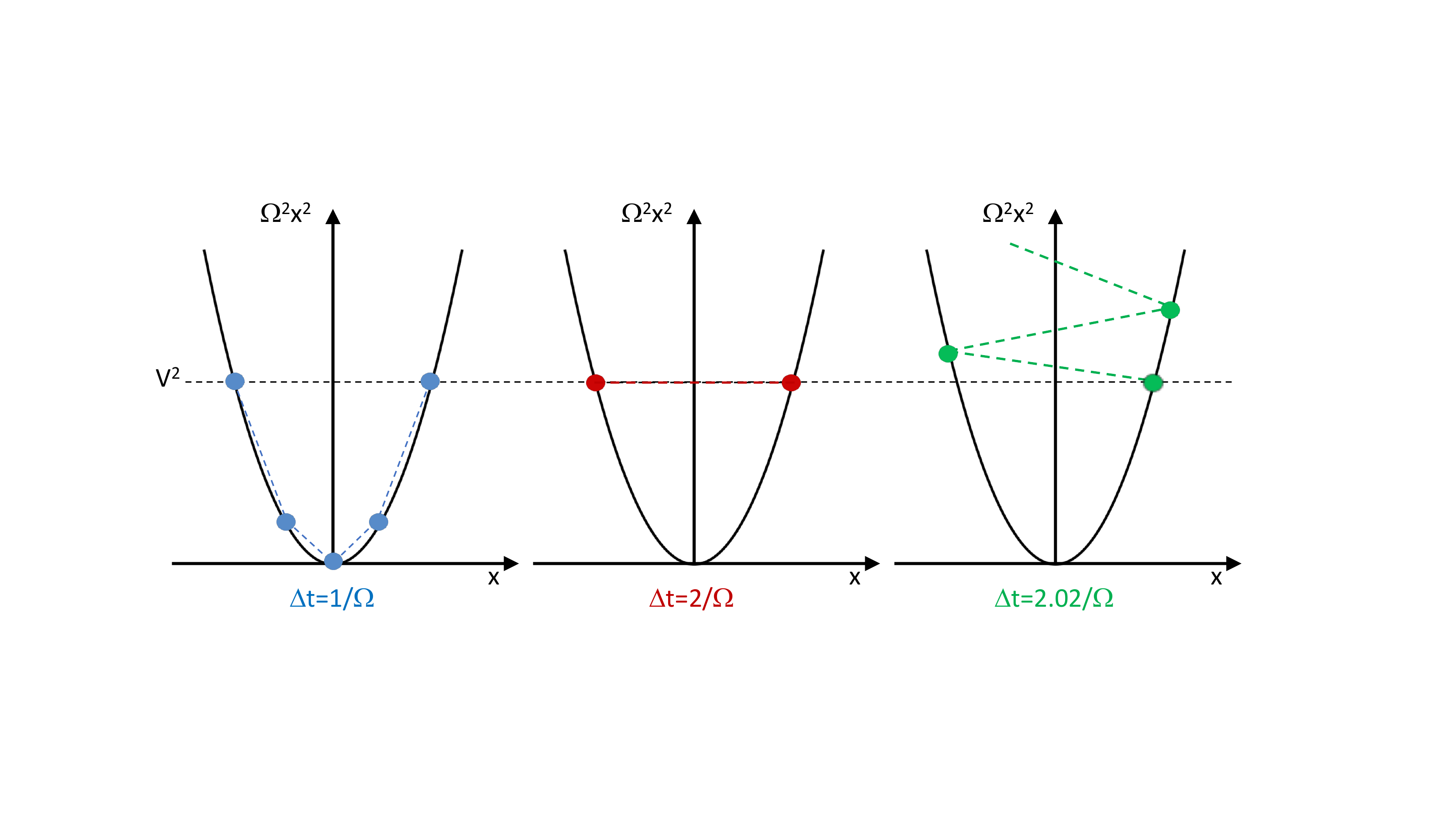}
\caption{\label{fig:harmonic_osc} Energy plots of second-order 
 finite-difference "leapfrog" integrator of the harmonic oscillator for three time steps:
 (left) $\Delta t = 1/\Omega$, (middle) $\Delta t = 2/\Omega$, and (right) $\Delta t = 2.02/\Omega$.}
\end{center}
\end{figure}
The origin of the CFL is easy to understand more intuitively by looking at the plot of the kinetic energy $\Omega^2 \mathbf{x}^2$ versus position for a harmonic oscillator initialized at rest with initial potential energy $\Omega^2 \mathbf{x}^2 = V^2$. The evolution of the kinetic energy is shown in Fig. \ref{fig:harmonic_osc} for a time step below the CFL ($\Delta t = 1/\Omega < 2/\Omega$), a time step at the CFL ($\Delta t = 2/\Omega$) and a time step above the CFL ( $\Delta t = 2.02/\Omega > 2/\Omega$).

\subsubsubsection{Numerical heating}

Due to various approximations, it is common to introduce errors that can be interpreted as a random forces $\delta F$ that give random velocity kicks $\delta v=\delta F \Delta t / m$. The average change in kinetic energy from these random kicks during one time step is given by
\begin{subequations}
\begin{eqnarray}
<h> & = & \frac{1}{2} m < | v_0 + \delta v|^2 > - \frac{1}{2} m < | v_0 |^2 >, \\
       & = & \frac{1}{2} m  (2 v_0 \cdot <\delta v > ) + \frac{1}{2} m < |\delta v|^2 >, \\
       & = & \frac{1}{2} m < |\delta v|^2 >,
\end{eqnarray}
\end{subequations}
since  $<\delta v >=0$.
Hence, the average energy change after $n$ time steps is given by
\begin{equation}
  E^n = E^0 + n \Delta t^2 \frac{|\delta F|^2}{2m}.
\end{equation}
This result shows that even if the numerical errors that affect the motion of the particles are random and average to zero, they will still result in a numerical ``stochastic heating'' that grows linearly in time and is proportional to the square of the time step. Since this type of error is typically unavoidable, it is important to always have it in mind when performing computer simulations.

\subsubsection{Full particle integrator}

The extension of the leapfrog integrator to the full set of relativistic Newton-Lorentz equations of motion is given by 
\begin{subequations}
\begin{align}
\frac{\mathbf{x}^{i+1}-\mathbf{x}^{i}}{\Delta t}= & \mathbf{v}^{i+1/2},\label{Eq:leapfrog_x}\\
\frac{\gamma^{i+1/2}\mathbf{v}^{i+1/2}-\gamma^{i-1/2}\mathbf{v}^{i-1/2}}{\Delta t}= & \frac{q}{m}\left(\mathbf{E}^{i}+\mathbf{\bar{v}}^{i}\times\mathbf{B}^{i}\right).\label{Eq:leapfrog_v}
\end{align}
\end{subequations}
Here again, all the quantities are centered properly to result in second-order centered finite-differences. 
This forces the velocity term that is involved in the magnetic rotation to be expressed as some average 
term $\bar{\mathbf{v}}^{i}$ to be determined.
In order to close the system, $\bar{\mathbf{v}}^{i}$ must be
expressed as a function of the other quantities. 
Some of the most popular implementations are presented below.

\subsubsubsection{Boris pusher}

The solution proposed by Boris \cite{BorisICNSP70} is given by 
\begin{align}
\mathbf{\bar{v}}^{i}= & \frac{\gamma^{i+1/2}\mathbf{v}^{i+1/2}+\gamma^{i-1/2}\mathbf{v}^{i-1/2}}{2\bar{\gamma}^{i}}.\label{Eq:boris_v}
\end{align}
where $\bar{\gamma}^{i}$ is defined by $\bar{\gamma}^{i} \equiv (\gamma^{i+1/2}+\gamma^{i-1/2} )/2$.

Setting $\mathbf{u}=\gamma\mathbf{v}$, the system of equations can be written as
\begin{subequations}
\begin{align}
\mathbf{u^{-}}= & \mathbf{u}^{i-1/2}+\left(q\Delta t/2m\right)\mathbf{E}^{i}  & \;\;\; \text{    (half acceleration)},\\
\mathbf{u}^{+} -\mathbf{u}^{-} = & \left(q\Delta t/2m\bar{\gamma^{i}}\right) (\mathbf{u}^{+} +\mathbf{u}^{-}) & \;\;\; \text{    (rotation)}, \\
\mathbf{u}^{i+1/2}= & \mathbf{u}^{+}+\left(q\Delta t/2m\right)\mathbf{E}^{i}  &\;\;\; \text{    (half acceleration)}.
\end{align}
\end{subequations}

Hence, the push is separated into one-half acceleration with the electric field, 
one rotation with the magnetic field and a second half acceleration with the electric field.
The rotation is solved very efficiently by the Boris' method using the following sequence:
\begin{subequations}
\begin{align}
\mathbf{u'}= & \mathbf{u}^{-}+\mathbf{u}^{-}\times\mathbf{t}\\
\mathbf{u}^{+}= & \mathbf{u}^{-}+\mathbf{u'}\times2\mathbf{t}/(1+t^{2})\\
\end{align}
\end{subequations}
where $\mathbf{t}=\left(q\Delta
  t/2m\right)\mathbf{B}^{i}/\bar{\gamma}^{i}$ and where
$\bar{\gamma}^{i}$ is given by $\bar{\gamma}^{i}=\sqrt{1+(\mathbf{u}^-/c)^2}$. 

The Boris implementation is second-order accurate, time-reversible and fast. Its implementation is very widespread and used in the vast majority of PIC codes.

\subsubsubsection{Vay pusher}

It was shown in \cite{VayPOP2008} that the Boris formulation does 
not capture properly the case of complete - of near-complete - cancellation 
of the electric acceleration and magnetic rotation, compromising for example
the modeling of ultra-relativistic charged particle beams.

An alternate velocity average was thus considered
\begin{align}
\mathbf{\bar{v}}^{i}= & \frac{\mathbf{v}^{i+1/2}+\mathbf{v}^{i-1/2}}{2},\label{Eq:new_v}
\end{align}
which was shown to treat accurately the complete - of near-complete - cancellation 
of the electric acceleration and magnetic rotation.

The proposed velocity average leads to a system that is solvable analytically (see \cite{VayPOP2008}
for a detailed derivation), giving the following velocity update:
\begin{subequations}
\begin{align}
\mathbf{u^{*}}= & \mathbf{u}^{i-1/2}+\frac{q\Delta t}{m}\left(\mathbf{E}^{i}+\frac{\mathbf{v}^{i-1/2}}{2}\times\mathbf{B}^{i}\right),\label{pusher_gamma}\\
\mathbf{u}^{i+1/2}= & \left[\mathbf{u^{*}}+\left(\mathbf{u^{*}}\cdot\mathbf{t}\right)\mathbf{t}+\mathbf{u^{*}}\times\mathbf{t}\right]/\left(1+t^{2}\right),\label{pusher_upr}
\end{align}
\end{subequations}
where $\mathbf{t}=\bm{\tau}/\gamma^{i+1/2}$, $\bm{\tau}=\left(q\Delta t/2m\right)\mathbf{B}^{i}$,
$\gamma^{i+1/2}=\sqrt{\sigma+\sqrt{\sigma^{2}+\left(\tau^{2}+w^{2}\right)}}$,
$w=\mathbf{u^{*}}\cdot\bm{\tau}$, $\sigma=\left(\gamma'^{2}-\tau^{2}\right)/2$
and $\gamma'=\sqrt{1+(\mathbf{u}^{*}/c)^{2}}$. 

\subsubsubsection{Higuera-Cary pusher}

The solution proposed by Higuera and Cary \cite{Higuera2017} adopted the same form 
for the velocity average as Boris
\begin{align}
\mathbf{\bar{v}}^{i}= & \frac{\gamma^{i+1/2}\mathbf{v}^{i+1/2}+\gamma^{i-1/2}\mathbf{v}^{i-1/2}}{2\bar{\gamma}^{i}},
\end{align}
but $\bar{\gamma}^{i}$ is now given by 
\begin{align}
\bar{\gamma}^{i} & \equiv \sqrt{1+\left(\frac{u^{i+1/2}+u^{i-1/2}}{2c} \right)^2}.
\end{align}
Similarly to the Boris, pusher, we have
\begin{subequations}
\begin{align}
\mathbf{u^{-}}= & \mathbf{u}^{i-1/2}+\left(q\Delta t/2m\right)\mathbf{E}^{i}  & \;\;\; \text{    (half acceleration)},\\
\mathbf{u}^{+}= & \mathbf{u}^{-}+(\mathbf{u}^{-}+\mathbf{u}^{-}\times\mathbf{t})\times2\mathbf{t}/(1+t^{2}) & \;\;\; \text{    (rotation)}, \\
\mathbf{u}^{i+1/2}= & \mathbf{u}^{+}+\left(q\Delta t/2m\right)\mathbf{E}^{i}  &\;\;\; \text{    (half acceleration)},
\end{align}
\end{subequations}
with $\mathbf{t}=\left(q\Delta t/2m\right)\mathbf{B}^{i}/\bar{\gamma}^{i}$, and where
\begin{align}
\bar{\gamma}^{i} = \sqrt{\frac{\sigma+\sqrt{\sigma^{2}+4\left(\tau^{2}+(u^{*})^2\right)}}{2}},
\end{align}
with 
$u^{*}=\mathbf{u^{-}}\cdot\bm{\tau}/c$, 
$\bm{\tau}=\left(q\Delta t/2m\right)\mathbf{B}^{i}$, 
$\sigma=\left((\gamma^{-})^{2}-\tau^{2}\right)/2$
and $\gamma^{-}=\sqrt{1+(\mathbf{u}^{-}/c)^{2}}$.

Like the Boris pusher, the Higuera-Cary pusher is strictly ``volume-preserving'', while the 
Vay pusher is not. Some comparison of the various pushers have been reported but 
there is not yet an exhaustive comparison of the three pushers for the modeling of plasma 
acceleration. This is an active area of research.

\subsection{Field solve}

Various methods are available for solving Maxwell's equations on a
grid, based on finite-differences, finite-volume, finite-element,
spectral, or other discretization techniques that apply most commonly
on single structured or unstructured meshes, and less commonly on multiblock
multiresolution grid structures. In this chapter, we summarize the widespread
second-order finite-difference time-domain (FDTD) algorithm, its extension
to non-standard finite-differences, as well as the pseudo-spectral
analytical time-domain (PSATD) algorithm. 

\subsubsection{Simple one-dimensional Leapfrog Maxwell solver}

\begin{figure}[ht]
\begin{center}
\includegraphics[trim={6.cm 3.5cm 5cm 3.cm},clip,scale=0.6]{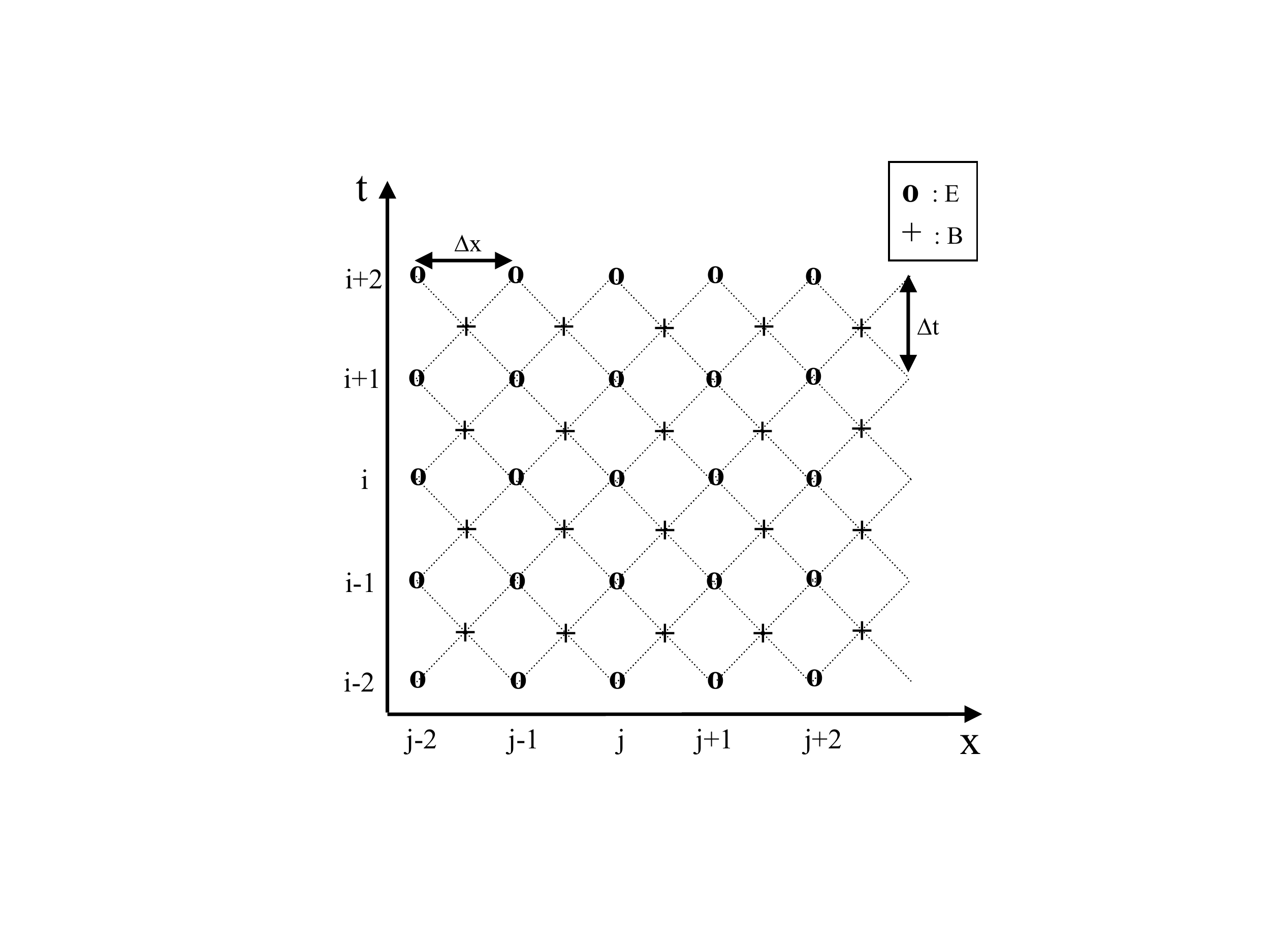}
\caption{\label{fig:maxwell1d} Space-time diagram illustrating the staggering of the 
field quantities in space and time of a one-dimensional, second-order, leapfrog 
wave integrator. The field E (circles) is known at integer positions $j+n$ and 
at integer time steps $i+m$, where $i$, $j$, $m$ and $n$ are integers. 
The field B (crosses) is known at half-integer positions $j+n+1/2$ and 
at half-integer time steps $i+m+1/2$. }
\end{center}
\end{figure}

For simplicity, we consider first a one-dimensional wave equation without source term:
\begin{subequations}
\begin{align}
\frac{\partial B}{\partial t}  = & c \frac{\partial E}{\partial x}, \\
\frac{\partial E}{\partial t}  = & c \frac{\partial B}{\partial x}.
\end{align}
\end{subequations}
Similarly to the particle pusher, this system of equations can easily be discretized 
using Leapfrog space- and time-centered finite-differencing, giving
\begin{subequations}
\begin{align}
\frac{B^{i+1/2}_{j+1/2}-B^{i-1/2}_{j+1/2}}{c\Delta t} = & \frac{E^{i}_{j+1}-E^{i}_{j}}{\Delta x}, \label{Eq:wave_1d_1}\\
\frac{E^{i+1}_{j}-E^{i}_{j}}{c\Delta t} = & \frac{B^{i+1/2}_{j+1/2}-B^{i+1/2}_{j-1/2}}{\Delta x}. \label{Eq:wave_1d_2}\
\end{align}
\end{subequations}

The electric and magnetic field components are defined on space-time grids that are staggered 
by half a space grid cell and half a time step, as depicted on Fig. \ref{fig:maxwell1d}.

\subsubsubsection{Numerical accuracy and stability analyses}

Performing a Von Neumann stability analysis, one makes the ansatz $E(x,t) = \mp B(x,t) = A e^{\i(\omega t \mp k x)}$ for 
waves propagating in the forward or backward direction. 
Applying to the terms in Eq. \ref{Eq:wave_1d_2} at $ i = j = 0$ gives

\begin{subequations}
\begin{align}
E^{i+1}_{j} = & A e^{\i(\omega \Delta t)}, \\
E^{i}_{j} = & A, \\
B^{i+1/2}_{j+1/2} = & A e^{\i(\omega \Delta t/2 \mp k \Delta x/2)}, \\
B^{i+1/2}_{j-1/2} = & A e^{\i(\omega \Delta t/2 \pm k \Delta x/2)}.
\end{align}
\end{subequations}

Inserting into Eq. \ref{Eq:wave_1d_2} and simplifying leads to
\begin{align}
\frac{\sin(\omega \Delta t/2)}{c\Delta t}= & \pm \frac{\sin(k\Delta x/2)}{\Delta x}. 
\end{align}

The phase and group velocities are thus given by
\begin{subequations}
\begin{align}
V_\phi = & \frac{\omega}{k} = \frac{2}{k\Delta t} \arcsin [ \frac{c\Delta t}{\Delta x} \sin (k\Delta x/2) ], \\
V_g = & \frac{\partial \omega}{\partial k} = \frac{c \cos(k\Delta x /2)}{\sqrt{1-[ \frac{c\Delta t}{\Delta x} \sin (k\Delta x/2) ]^2}}.
\end{align}
\end{subequations}

\begin{figure}[ht]
\begin{center}
\includegraphics[trim={0.cm 7.cm 4cm 0.cm},clip,scale=0.5]{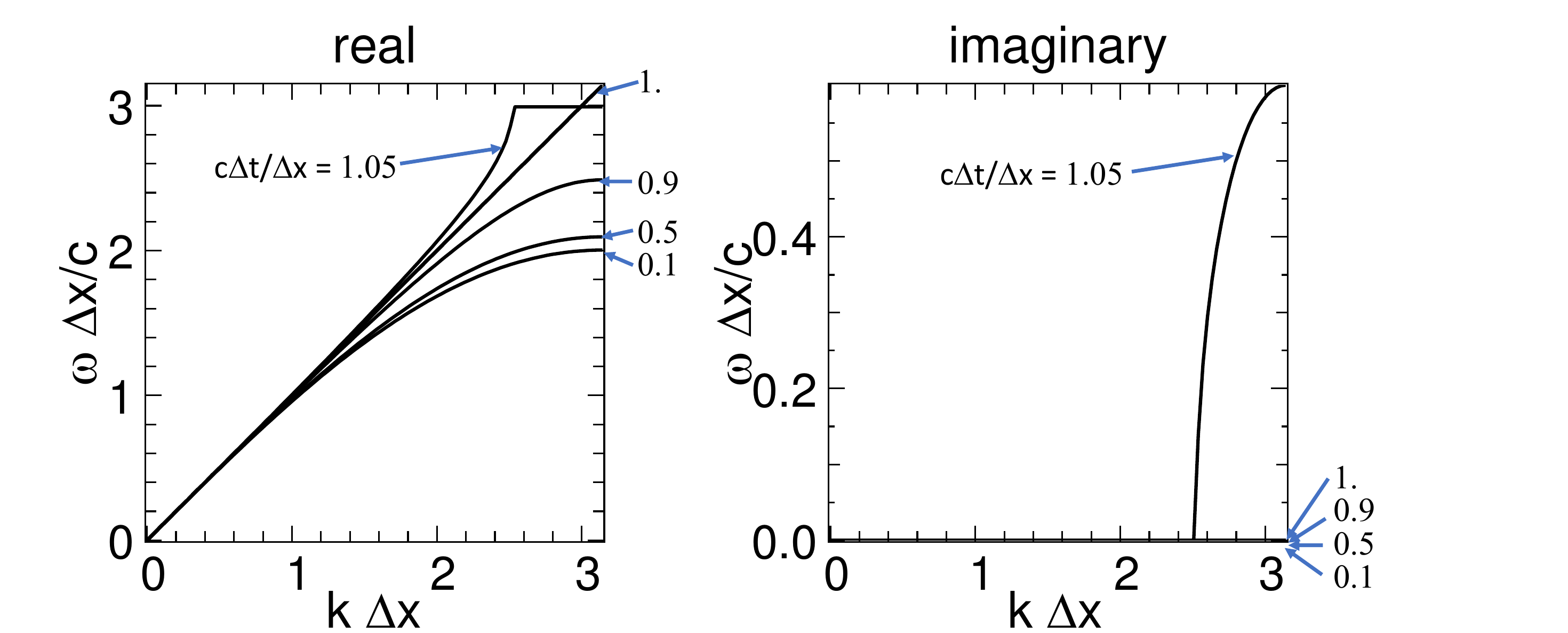}
\includegraphics[trim={0.cm 7.cm 4cm 0.cm},clip,scale=0.5]{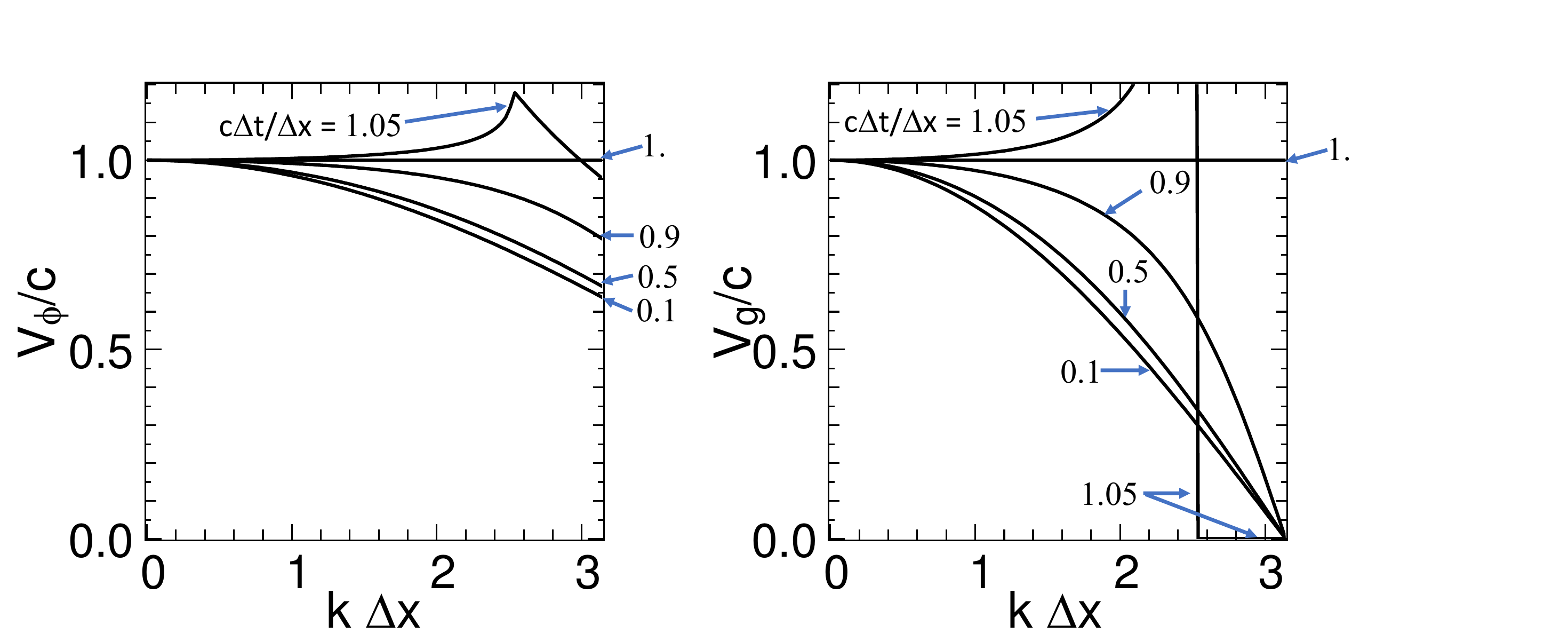}
\caption{\label{fig:disp1d} (top) Real part (left) and imaginary part (right) of the 
numerical dispersion of the one-dimensional, second-order, leapfrog 
wave equation integrator for $c\Delta t/\Delta x=\{0.1,0.5,0.9,1.,1.05\}$. 
(bottom) Corresponding phase velocity (left) and group velocity (right).}
\end{center}
\end{figure}

The values of $\omega$, $V_\phi$ and $V_g$ are plotted in Fig. \ref{fig:disp1d} 
as a function of $k$, for various values of $c\Delta t/\Delta x$, 
leading to the following observations:
\begin{itemize}
\item $\bf c\Delta t/\Delta x < 1$: the imaginary part of $\omega$ is null and the 
scheme is stable. The phase  and group velocities are below the physical value $c$, 
respectively as low as $~2/3 c$ and $0$ at the Nyquist cutoff. 
\item $\bf c\Delta t/\Delta x=1$: the physical result 
$\omega=kc$ is recovered, and all quantities are exact at all wavelengths. The 
time step $\Delta t = \Delta x/c$ is sometimes referred to as the ``magical time step''.
\item $\bf c\Delta t/\Delta x>1$: the imaginary part is positive, meaning that the 
solution is unstable and grows exponentially in time.
\end{itemize}

\begin{figure}[ht]
\begin{center}
\includegraphics[trim={0.cm 3.cm 2.5cm 0.cm},clip,scale=0.5]{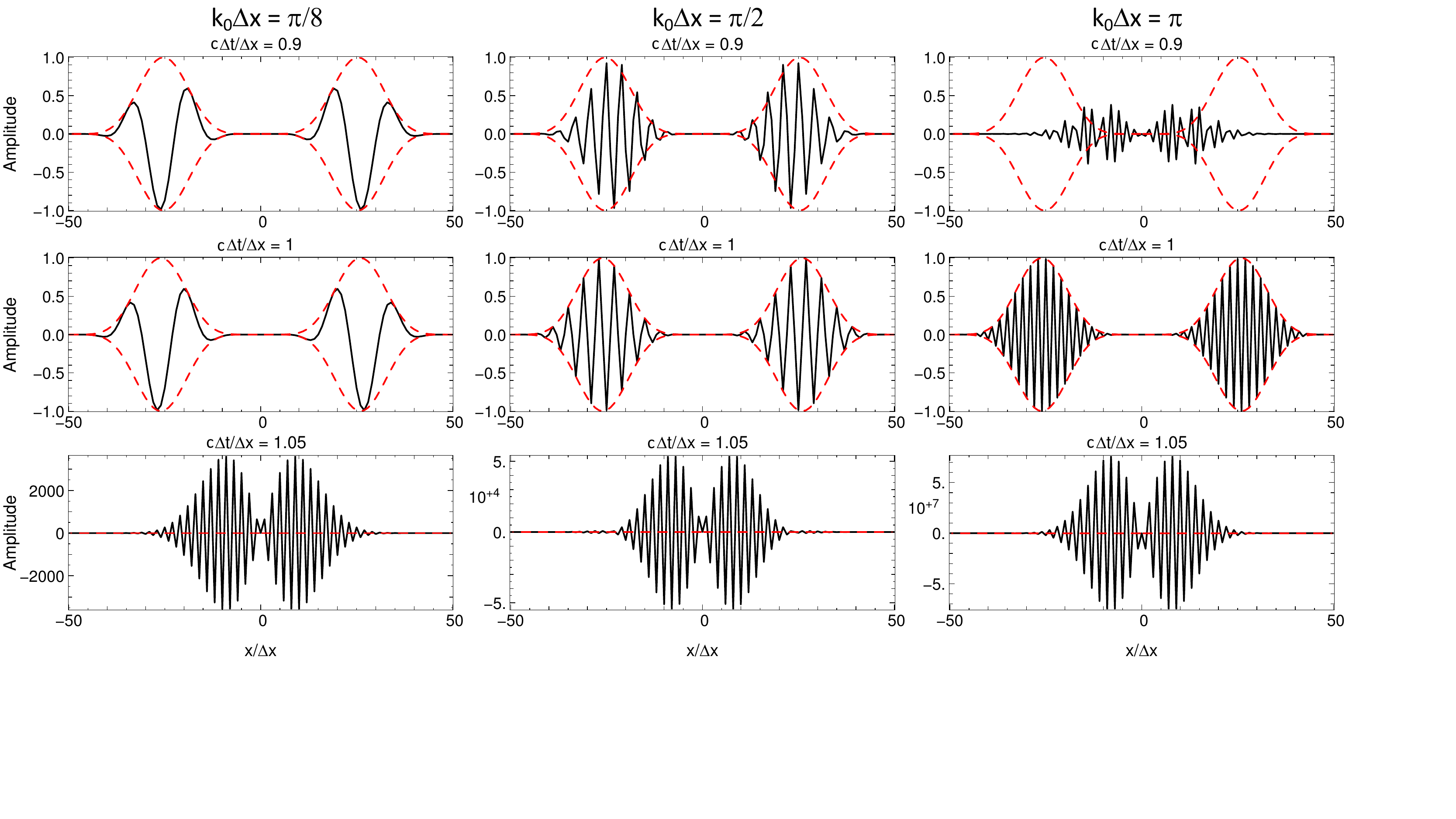}
\caption{\label{fig:FDTD_scan_j0} Snapshots from simulations using the one-dimensional, second-order, leapfrog 
wave equation integrator for (left) $k_0 \Delta x = \pi / 8$, (center)  $k_0 \Delta x = \pi / 2$, 
(right) $k_0 \Delta x = \pi$, with (top) $c\Delta t/\Delta x = 0.9$, (middle) $c\Delta t/\Delta x = 1$ 
and (bottom) $c\Delta t/\Delta x = 1.05$. A driving signal is imposed on the field $E$ at 
the central grid point of the simulation grid, generating two pulses (propagating respectively 
forward and backward) composed of a wave at a single frequency and modulated by an envelope (given by a Harris 
function). The field $E$ is plotted (solid black line) after $N c \Delta x / (2\Delta t)$ time steps. The 
analytical solution of the propagation of the envelope is plotted (dashed red lines) for reference.}
\end{center}
\end{figure}

Snapshots from simulations using the Maxwell solver for various wavelengths and 
time steps are shown in Fig. \ref{fig:FDTD_scan_j0}, illustrating the findings of the numerical 
analysis. A driving signal is imposed on the field $E$ at the central grid point of the simulation grid, 
of the form $E(t) = H(t) \cos(\omega_0 t)$ where $H$ is the Harris function 
$H(t) = (10-15\cos(2\pi t/T)+6\cos(4\pi t/T)-\cos(6\pi t/T))/32$ for $0<t<T$ and $0$ otherwise, with 
$T = N \Delta x / (2c)$, $\omega_0 = 2/\Delta t \arcsin(\min(1,\Delta t \sin(k_0/2)))$ and where 
$N$ is the number of grid cells. Plots are given for  $k_0 \Delta x = \{\pi/8, \pi/2, \pi\}$ and 
$c\Delta t/\Delta x = \{0.9,1.,1.05\}$.

For $c\Delta t/\Delta x = 0.9$, the accuracy of the wave propagation deteriorates with increasing $\omega_0$, 
to the point of being very inaccurate at the Nyquist limit. At $c\Delta t/\Delta x = 1$, the pulses are 
correctly propagated, with the correct phase and amplitude, as expected. With $c\Delta t/\Delta x = 1.05$, 
the instability of the solver is clearly visible, with the largest growth rate at the Nyquist wavelength $k_0 \Delta x = \pi$.

\subsubsubsection{One-dimensional wave solver with source term}

While the analysis and the numerical experiment shows that the algorithm is stable and accurate 
for $c\Delta t/\Delta = 1$, it is illuminating to look at the response of the algorithm to a 
prescribed source term. 

The wave equation with source term writes as:
\begin{subequations}
\begin{align}
\frac{\partial B}{\partial t}  = & c \frac{\partial E}{\partial x}, \\
\frac{\partial E}{\partial t}  = & c \frac{\partial B}{\partial x}-J.
\end{align}
\end{subequations}

Similarly to the particle pusher, this system of equations can easily be discretized 
using leapfrogged space- and time-centered finite-differencing, giving
\begin{subequations}
\begin{align}
\frac{B^{i+1/2}_{j+1/2}-B^{i-1/2}_{j+1/2}}{\Delta t} = & \frac{E^{i}_{j+1}-E^{i}_{j}}{\Delta x}, \label{fdtd1dwithJ1} \\ 
\frac{E^{i+1}_{j}-E^{i}_{j}}{\Delta t} = & \frac{B^{i+1/2}_{j+1/2}-B^{i+1/2}_{j-1/2}}{\Delta x} - J^{i+1/2}_j. \label{fdtd1dwithJ2}
\end{align}
\end{subequations}
\begin{figure}[ht]
\begin{center}
\includegraphics[trim={0.cm 3.cm 2.5cm 0.cm},clip,scale=0.5]{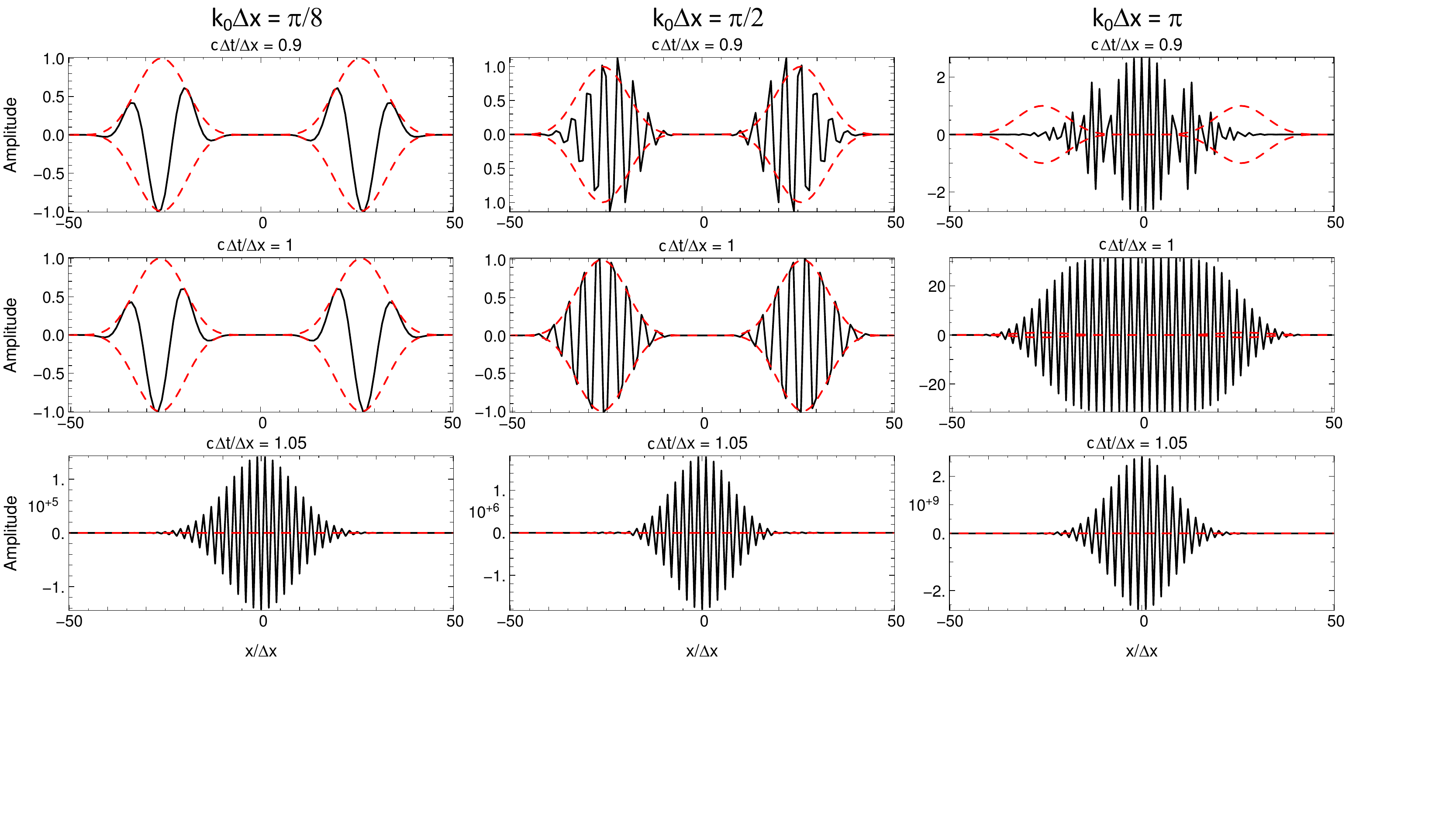}
\caption{\label{fig:FDTD_scan_j1} Snapshots from simulations using the one-dimensional, second-order, leapfrog 
wave equation integrator for (left) $k \Delta x = \pi / 8$, (center)  $k \Delta x = \pi / 2$, 
(right) $k \Delta x = \pi$, with (top) $c\Delta t/\Delta x = 0.9$, (middle) $c\Delta t/\Delta x = 1.$ 
and (bottom) $c\Delta t/\Delta x = 1.05$.}
\end{center}
\end{figure}

Snapshots from simulations using the Maxwell solver with a prescribed source term, driven at various wavelengths and 
time steps are shown in Fig. \ref{fig:FDTD_scan_j1}. The most notable differences with the scheme without source 
is that the signal grows to larger amplitudes at the Nyquist wavelength, especially for $c\Delta t/\Delta x = 1$. 
Indeed, analysis shows that the response of Eq. \ref{fdtd1dwithJ1}-\ref{fdtd1dwithJ2} to a source oscillating at 
$k_0 \Delta x = \pi$ is of the form $E(n) = -n \times (-1)^n$, exhibiting a linear amplitude growth in time.
This indicates that when coupling with particles, it is important to remove any signal at the Nyquist wavelength. 
This can be done efficiently by applying a bilinear filter (see section \ref{Sec:filtering}) on the source term at each time step.

\subsubsection{The Yee or Finite-Difference Time-Domain (FDTD) Maxwell solver}

\begin{figure}[ht]
\begin{center}
\includegraphics[trim={1.cm 1.2cm 1cm 10cm},clip,scale=0.6]{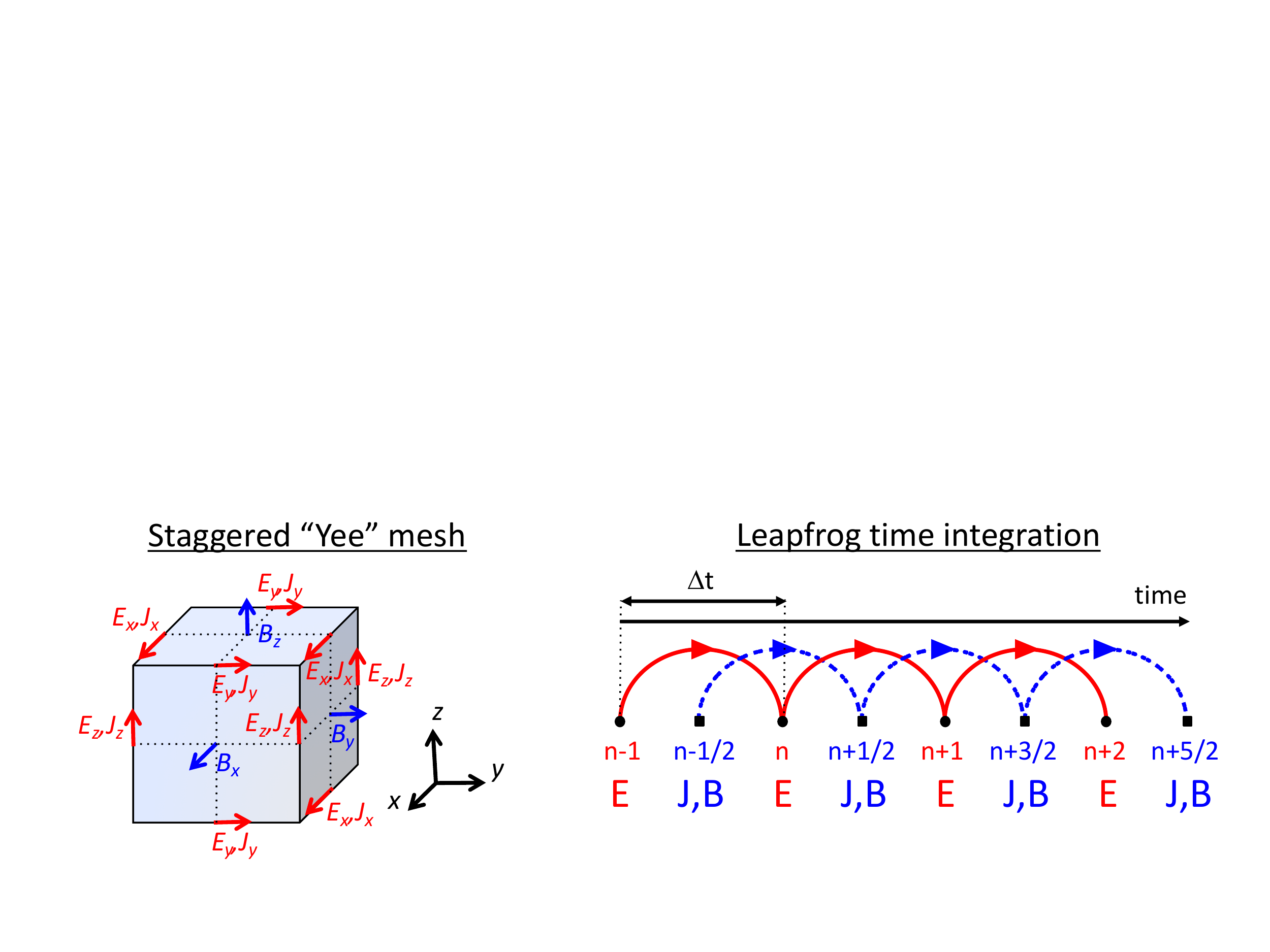}
\caption{\label{fig:yee_grid}(left) Layout of field components on the staggered ``Yee''
grid. Current densities and electric fields are defined on the edges of the cells and
 magnetic fields on the faces. (right) Time integration using a second-order 
 finite-difference "leapfrog" integrator.}
\end{center}
\end{figure}

The most popular algorithm for electromagnetic PIC codes is the Yee, 
also known as Finite-Difference Time-Domain (or FDTD), solver \cite{Yee}.
It defines the electromagnetic field components on a staggered grid, such 
that all the $\nabla\times$ operations in the discretized Maxwell's equations are 
properly centered. The layout of the components is given in Fig. \ref{fig:yee_grid}, 
where the electric field components are located
between nodes and the magnetic field components are located in the
center of the cell faces. The time integration follows the leapfrop scheme, where, 
knowing the current densities at half-integer steps, 
the electric field components are updated alternately with the magnetic 
field components at integer and half-integer steps respectively.

The algorithm can be written
\begin{subequations}
\begin{eqnarray}
D_{t}\mathbf{B} & = & -\nabla\times\mathbf{E}\label{Eq:Faraday-2},\\
D_{t}\mathbf{E} & = & \nabla\times\mathbf{B}-\mathbf{J}\label{Eq:Ampere-2}, 
\end{eqnarray}
\end{subequations}
with the following definitions of the operators:
\begin{subequations}
\begin{eqnarray}
D_{t}G|_{i,j,k}^{n}= & \left(G|_{i,j,k}^{n+1/2}-G|_{i,j,k}^{n-1/2}\right)/\Delta t, \\
D_{x}G|_{i,j,k}^{n}= & \left(G|_{i+1/2,j,k}^{n}-G|_{i-1/2,j,k}^{n}\right)/\Delta x, \\
D_{y}G|_{i,j,k}^{n}= & \left(G|_{i,j+1/2,k}^{n}-G|_{i,j-1/2,k}^{n}\right)/\Delta y, \\
D_{z}G|_{i,j,k}^{n}= & \left(G|_{i,j,k+1/2}^{n}-G|_{i,j,k-1/2}^{n}\right)/\Delta z, \\
\nabla= & D_{x}\mathbf{\hat{x}}+D_{y}\mathbf{\hat{y}}+D_{z}\mathbf{\hat{z}},
\end{eqnarray}
\end{subequations}
where $\Delta t$, $\Delta x$, $\Delta y$ and $\Delta z$ are respectively the time step and
the grid cell sizes along $x$, $y$ and $z$, respectively, and where 
$n$ is the time index and $i$, $j$
and $k$ are the spatial indices along $x$, $y$ and $z$ respectively.

For example, the update of $E_x$ is given explicitly by:
\begin{align}
\frac{E_x\rvert_{i+1/2,j,k}^{n+1} - E_x\rvert_{i+1/2,j,k}^n}{\Delta t} = & \frac{B_z\rvert_{i+1/2,j+1/2,k}^{n+1/2} - B_z\rvert_{i+1/2,j-1/2,k}^{n+1/2}}{\Delta y} \\
- & \frac{B_y\rvert_{i+1/2,j,k+1/2}^{n+1/2} - B_y\rvert_{i+1/2,j,k-1/2}^{n+1/2}}{\Delta z} \\
- & J_x\rvert_{i+1/2,j,k}^{n+1/2}.
\end{align}
The updates for $E_y$, $E_z$, $B_x$, $B_y$ and $B_z$ are obtained similarly.

\subsubsubsection{Numerical accuracy and stability analysis}
A Von Neumann analysis leads to the following relation of dispersion
\begin{subequations}
\begin{eqnarray}
\left[ \frac{\sin(\omega \Delta t/2)}{c\Delta t} \right]^2 & = & \left[ \frac{\sin(k_x\Delta x/2)}{\Delta x} \right]^2 
                                                                                      + \left[ \frac{\sin(k_y\Delta y/2)}{\Delta y} \right]^2, \\
\left[ \frac{\sin(\omega \Delta t/2)}{c\Delta t} \right]^2 & = & \left[ \frac{\sin(k_x\Delta x/2)}{\Delta x} \right]^2 
                                                                                      + \left[ \frac{\sin(k_y\Delta y/2)}{\Delta y} \right]^2 
                                                                                      + \left[ \frac{\sin(k_z\Delta z/2)}{\Delta z} \right]^2, 
\end{eqnarray}
\end{subequations}
in two and three dimensions, respectively. This results in the following CFL conditions:
\begin{subequations}
\begin{eqnarray}
c \Delta t \leq \frac{1}{\sqrt{\frac{1}{\Delta x^2}+\frac{1}{\Delta y^2}}}  & \text {in 2-D}, \\
c \Delta t \leq \frac{1}{\sqrt{\frac{1}{\Delta x^2}+\frac{1}{\Delta y^2}+\frac{1}{\Delta z^2}}} & \text {in 3-D}.
\end{eqnarray}
\end{subequations}

When $\Delta x=\Delta y$ and $\Delta x=\Delta y = \Delta z$, the CFL conditions reduce to $c\Delta t<\Delta x/\sqrt{2}$
and $c\Delta t<\Delta x/\sqrt{3}$ in 2D and 3D, respectively.

The phase and group velocities are given respectively by
\begin{subequations}
\begin{align}
V_\phi^\text{(2D)} = & \frac{\omega}{|\mathbf{k}|} = \frac{\arcsin \left[ c\Delta t \sqrt{ \left(\frac{\sin(k_x \Delta x/2)}{\Delta x}\right)^2 + \left(\frac{\sin(k_y \Delta y/2)}{\Delta y}\right)^2 } \right]}{|\mathbf{k}|\Delta t/2}, \\
V_g^\text{(2D)} = & \left|\frac{\text{d} \omega}{\text{d} \mathbf{k}} \right| = \frac{\sqrt{ \left(\frac{\sin(k_x \Delta x)}{\Delta x}\right)^2 
                                                                                                                  + \left(\frac{\sin(k_y \Delta y)}{\Delta y}\right)^2 } } {\frac{\sin(\omega \Delta t)}{c \Delta t }}
\end{align}
\end{subequations}
in 2-D, and 
\begin{subequations}
\begin{align}
V_\phi^\text{(3D)} = & \frac{\omega}{|\mathbf{k}|} = \frac{\arcsin \left[ c\Delta t \sqrt{ \left(\frac{\sin(k_x \Delta x/2)}{\Delta x}\right)^2 
                                            + \left(\frac{\sin(k_y \Delta y/2)}{\Delta y}\right)^2 
                                            + \left(\frac{\sin(k_z \Delta z/2)}{\Delta z}\right)^2 } \right]}{|\mathbf{k}|\Delta t/2}, \\
V_g^\text{(3D)} = & \left|\frac{\text{d} \omega}{\text{d} \mathbf{k}}\right|  =  \frac{\sqrt{ \left(\frac{\sin(k_x \Delta x)}{\Delta x}\right)^2 
                                                                                                 + \left(\frac{\sin(k_y \Delta y)}{\Delta y}\right)^2  
                                                                                                 + \left(\frac{\sin(k_y \Delta z)}{\Delta z}\right)^2 } } {\frac{\sin(\omega \Delta t)}{c \Delta t}}.
\end{align}
\end{subequations}
in 3-D.

\begin{figure}[ht]
\begin{center}
\includegraphics[trim={0.cm 7.cm 6cm 0.cm},clip,scale=0.5]{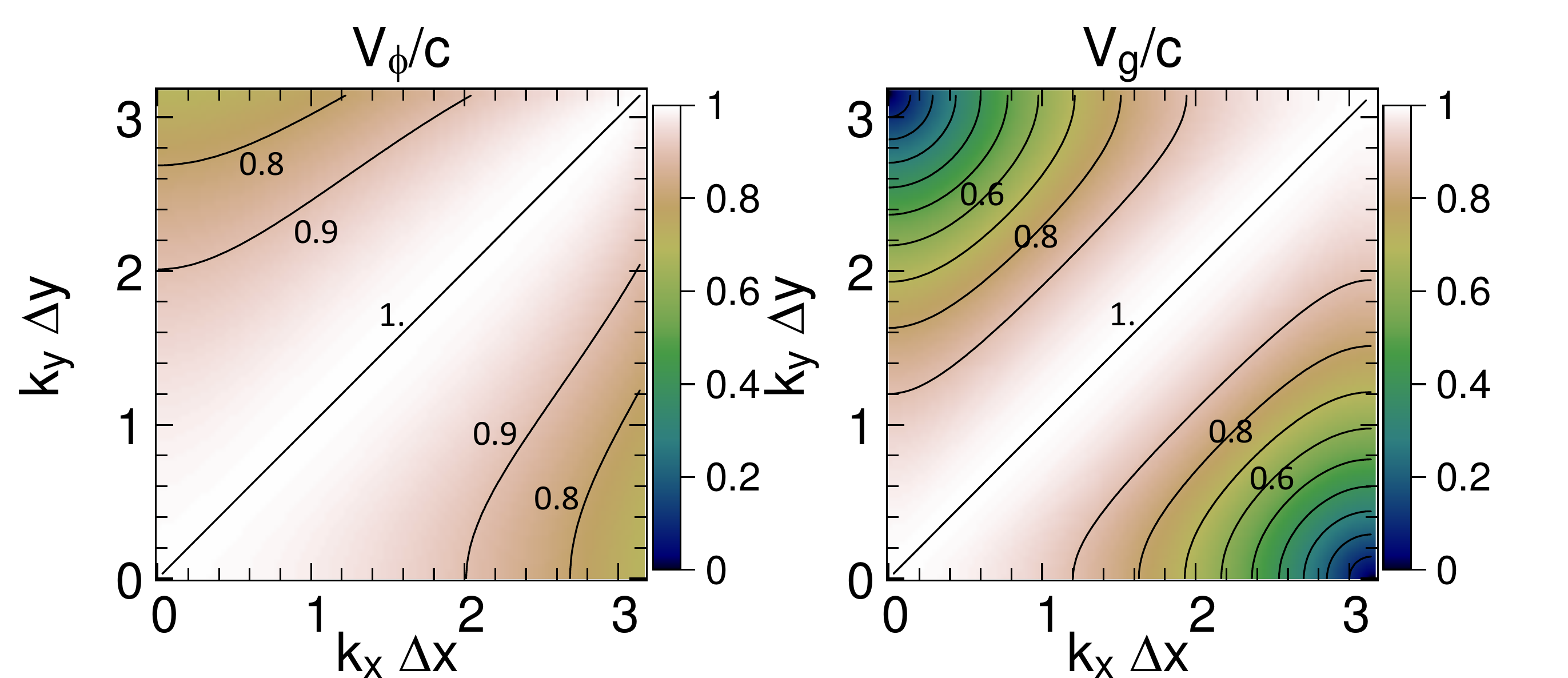}
\caption{\label{fig:Vyee2d} (left) Phase velocity and (right) group velocity 
from the 2-D Yee solver versus wavenumbers, when $\Delta y=\Delta x$ and 
$c \Delta t = \Delta x/\sqrt{2}$.}
\end{center}
\end{figure}

The maps of the 2-D phase and group velocities are plotted in Fig. \ref{fig:Vyee2d}, 
using the time step at the Courant limit $c\Delta t = 1/\sqrt{\frac{1}{\Delta x^2}+\frac{1}{\Delta y^2}}$, 
with $\Delta y=\Delta x$.
In this case, the phase and group velocities are exact for $k_y\Delta y=k_x\Delta x$, i.e. for a planar wave 
propagating at 45 degrees from the grid axes. At other angles, both the phase and group velocities are 
below the physical value, down to $c/\sqrt{2}$ and to $0$ for the phase and group velocities respectively,
at the Nyquist wavelength along the grid axes.

In 3-D, the phase and group velocities are exact along the main diagonal of the grid 
when using the time step at the Courant limit $c\Delta t = 1/\sqrt{\frac{1}{\Delta x^2}+\frac{1}{\Delta y^2}+\frac{1}{\Delta z^2}}$, 
with $\Delta z=\Delta y=\Delta x$. At other angles, both the phase and group velocities are 
below the physical value.

In the modeling of laser-driven plasma acceleration, it is convenient to launch the laser 
along one of the main axes, and the numerical dispersion of the Yee solver can lead 
to significant errors. To remedy this issue, it has become common to adopt solvers 
based on Non-Standard Finite-Difference Time-Domain (NSFDTD) or 
Pseudo-Spectral Analytical Time-Domain solvers, which will be presented next.


\subsubsection{Non-Standard Finite-Difference Time-Domain (NSFDTD)}
\begin{figure}[ht]
\begin{center}
\includegraphics[trim={0.cm 14.cm 16cm 0cm},clip,scale=1.]{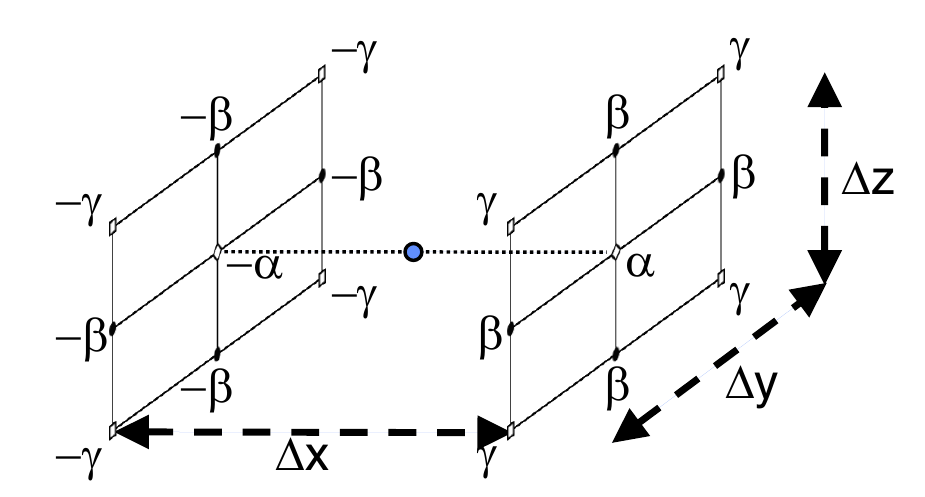}
\caption{\label{fig:NSFDTDoperator} Layout of the Non-Standard Finite-Difference 
Time-Domain (NSFDTD) operator. In this example, the finite difference is performed 
between two adjacent points along $x$, to evaluate the derivative at the central 
location (blue-filled circle). In addition to the points along x, points along the directions 
$y$ and $z$ are also taken into account. Coefficients $\alpha$, $\beta$ and $\gamma$ 
are assigned with the requirement that $\alpha+4\beta+4\gamma=1$.
}
\end{center}
\end{figure}

In \cite{ColeIEEE1997,ColeIEEE2002}, Cole introduced an implementation
of the source-free Maxwell's wave equations for narrow-band applications
based on non-standard finite-differences (NSFD). In \cite{KarkICAP06},
Karkkainen \emph{et al.} adapted it for wideband applications. At
the Courant limit for the time step and for a given set of parameters,
the stencil proposed in \cite{KarkICAP06} has no numerical dispersion
along the principal axes, provided that the cell size is the same
along each dimension (i.e. cubic cells in 3D). The ``Cole-Karkkainnen''
(or CK) solver uses the non-standard finite difference formulation
(based on extended stencils) of the Maxwell-Ampere equation. The 
extension for electromagnetic PIC with the source term is given 
in \cite{VayJCP2011} and reads:
\begin{subequations}
\begin{eqnarray}
D_{t}\mathbf{B} & = & -\nabla^{*}\times\mathbf{E}, \label{Eq:Faraday}\\
D_{t}\mathbf{E} & = & \nabla\times\mathbf{B}-\mathbf{J}. \label{Eq:Ampere}
\end{eqnarray}
\end{subequations}
%
The NSFD differential operator $\nabla^{*}$ (see layout in Fig. \ref{fig:NSFDTDoperator}) is given by 
\begin{align}
\nabla^{*}=D_{x}^{*}\mathbf{\hat{x}}+D_{y}^{*}\mathbf{\hat{y}}+D_{z}^{*}\mathbf{\hat{z}},
\end{align}
where 
\begin{align}
D_{x}^{*}=\left(\alpha+\beta S_{x}^{1}+\xi S_{x}^{2}\right)D_{x},
\end{align}
with 
\begin{subequations}
\begin{eqnarray}
S_{x}^{1}G|_{i,j,k}^{n} & = & G|_{i,j+1,k}^{n}+G|_{i,j-1,k}^{n}+G|_{i,j,k+1}^{n}+G|_{i,j,k-1}^{n}, \\
S_{x}^{2}G|_{i,j,k}^{n} & = & G|_{i,j+1,k+1}^{n}+G|_{i,j-1,k+1}^{n}+G|_{i,j+1,k-1}^{n}+G|_{i,j-1,k-1}^{n}.
\end{eqnarray}
\end{subequations}

$G$ is a sample vector component, while $\alpha$, $\beta$ and $\xi$
are constant scalars satisfying $\alpha+4\beta+4\xi=1$. As with
the FDTD algorithm, the quantities with half-integer are located between
the nodes (electric field components) or in the center of the cell
faces (magnetic field components). The operators along $y$ and $z$,
i.e. $D_{y}$, $D_{z}$, $D_{y}^{*}$, $D_{z}^{*}$, $S_{y}^{1}$,
$S_{z}^{1}$, $S_{y}^{2}$, and $S_{z}^{2}$, are obtained by circular
permutation of the indices.

For example, the update of $B_x$ is given explicitly by:
\begin{align}
\frac{B_x\rvert_{i,j+1/2,k+1/2}^{n+1/2} - B_x\rvert_{i,j+1/2,k+1/2}^{n-1/2}}{\Delta t}   = \nonumber  \\
\alpha \left[ \frac{E_y\rvert_{i,j+1/2,k+1}^{n} - E_y\rvert_{i,j+1/2,k}^{n}}{\Delta z} \right. 
                    - \left. \frac{E_z\rvert_{i,j+1,k+1/2}^{n} - E_z\rvert_{i,j,k+1/2}^{n}}{\Delta y}  \right] \nonumber \\
            +  \beta \left[ \frac{E_y\rvert_{i+1,j+1/2,k+1}^{n} - E_y\rvert_{i+1,j+1/2,k}^{n}}{\Delta z} \right.
                       - \left. \frac{E_z\rvert_{i+1,j+1,k+1/2}^{n} - E_z\rvert_{i+1,j,k+1/2}^{n}}{\Delta y} \right. \nonumber \\
                        + \left. \frac{E_y\rvert_{i-1,j+1/2,k+1}^{n} - E_y\rvert_{i-1,j+1/2,k}^{n}}{\Delta z} \right. 
                       - \left. \frac{E_z\rvert_{i-1,j+1,k+1/2}^{n} - E_z\rvert_{i-1,j,k+1/2}^{n}}{\Delta y} \right. \nonumber \\
                        + \left. \frac{E_y\rvert_{i,j+3/2,k+1}^{n} - E_y\rvert_{i,j+3/2,k}^{n}}{\Delta z} \right. 
                       - \left. \frac{E_z\rvert_{i,j+1,k+3/2}^{n} - E_z\rvert_{i,j,k+3/2}^{n}}{\Delta y} \right. \nonumber \\
                       + \left.  \frac{E_y\rvert_{i,j-1/2,k+1}^{n} - E_y\rvert_{i,j-1/2,k}^{n}}{\Delta z} \right. 
                        - \left. \frac{E_z\rvert_{i,j+1,k-1/2}^{n} - E_z\rvert_{i,j,k-1/2}^{n}}{\Delta y}  \right] \nonumber\\
            +  \gamma \left[ \frac{E_y\rvert_{i+1,j+3/2,k+1}^{n} - E_y\rvert_{i+1,j+3/2,k}^{n}}{\Delta z} \right.
                       - \left. \frac{E_z\rvert_{i+1,j+1,k+3/2}^{n} - E_z\rvert_{i+1,j,k+3/2}^{n}}{\Delta y} \right. \nonumber \\
                        + \left. \frac{E_y\rvert_{i-1,j+3/2,k+1}^{n} - E_y\rvert_{i-1,j+3/2,k}^{n}}{\Delta z} \right. 
                       - \left. \frac{E_z\rvert_{i-1,j+1,k+3/2}^{n} - E_z\rvert_{i-1,j,k+3/2}^{n}}{\Delta y} \right. \nonumber \\
                        + \left. \frac{E_y\rvert_{i+1,j-1/2,k+1}^{n} - E_y\rvert_{i+1,j-1/2,k}^{n}}{\Delta z} \right. 
                       - \left. \frac{E_z\rvert_{i+1,j+1,k-1/2}^{n} - E_z\rvert_{i+1,j,k-1/2}^{n}}{\Delta y} \right. \nonumber \\
                       + \left.  \frac{E_y\rvert_{i-1,j-1/2,k+1}^{n} - E_y\rvert_{i-1,j-1/2,k}^{n}}{\Delta z} \right. 
                        - \left. \frac{E_z\rvert_{i-1,j+1,k-1/2}^{n} - E_z\rvert_{i-1,j,k-1/2}^{n}}{\Delta y}  \right].
\end{align}

The updates of $B_y$ and $B_z$ are obtained by circular permutations. The updates of $E_x$, $E_y$ and $E_z$ 
are the same as for the Yee algorithm.


\subsubsubsection{Numerical accuracy and stability analysis}
A Von Neumann analysis leads to the following relation of dispersion
\begin{subequations}
\begin{eqnarray}
\left[ \frac{\sin(\omega \Delta t/2)}{c\Delta t} \right]^2 & = & C_x \left[ \frac{\sin(k_x\Delta x/2)}{\Delta x} \right]^2 
                                                                                      + C_y \left[ \frac{\sin(k_y\Delta y/2)}{\Delta y} \right]^2 
                                                                                      + C_z  \left[ \frac{\sin(k_z\Delta z/2)}{\Delta z} \right]^2, 
\end{eqnarray}
\end{subequations}
with 
\begin{subequations}
\begin{eqnarray}
C_x = & \alpha  + 2 \beta ( c_y + c_z) + 4 \gamma c_y c_z, \\
C_y = & \alpha  + 2 \beta ( c_x + c_z) + 4 \gamma c_x c_z, \\
C_z = & \alpha  + 2 \beta ( c_x + c_y) + 4 \gamma c_x c_y, \\
\end{eqnarray}
\end{subequations}
and
\begin{subequations}
\begin{eqnarray}
c_x = &\cos(k_x\Delta x), \\
c_y = &\cos(k_y\Delta y), \\
c_z = &\cos(k_z\Delta z). \\
\end{eqnarray}
\end{subequations}

As shown in \cite{KarkICAP06}, assuming $\Delta z=\Delta y=\Delta x$, and 
setting $\alpha=7/12$, $\beta=1/12$ and $\gamma=1/48$ leads to a 
CFL limit of $c\Delta t/\Delta x=1$.

\begin{figure}[ht]
\begin{center}
\includegraphics[trim={0.cm 7.cm 6cm 0.cm},clip,scale=0.5]{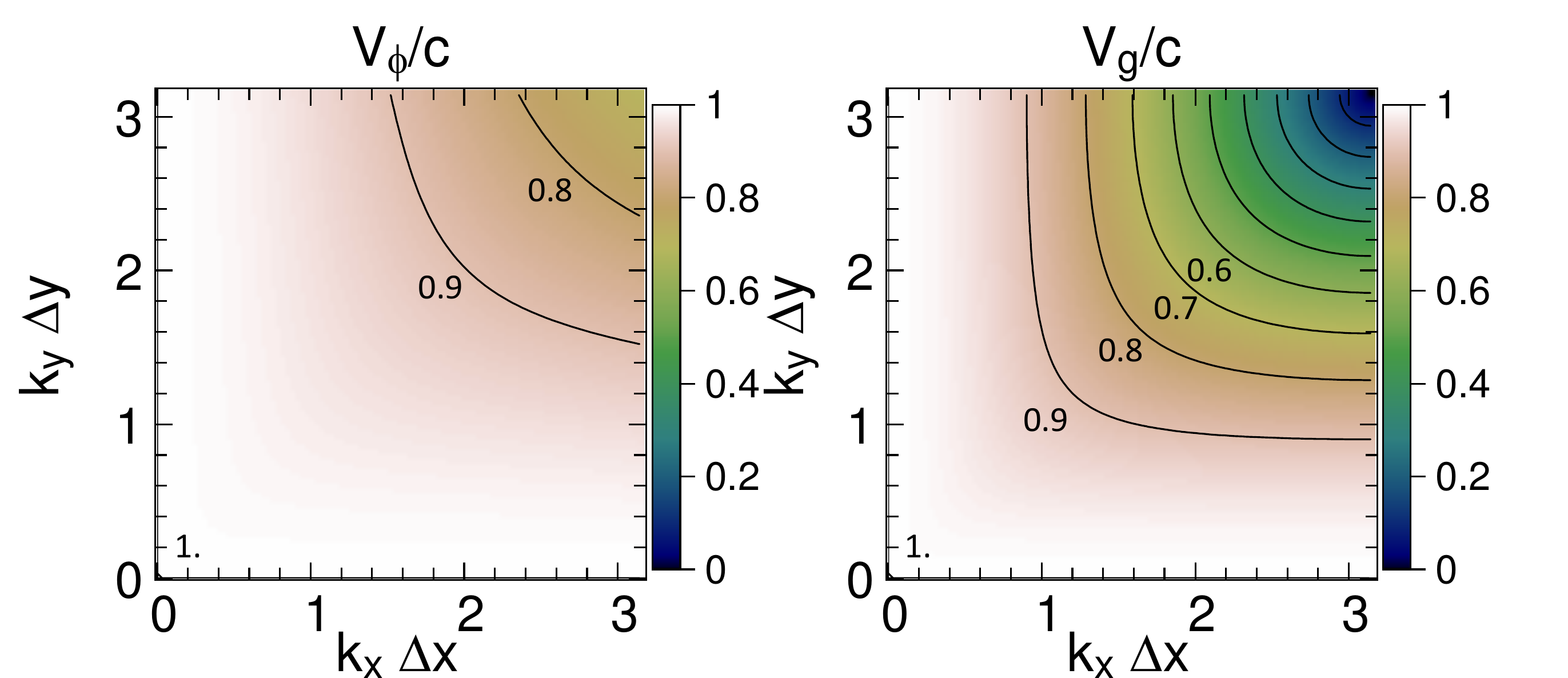}
\caption{\label{fig:Vckc2d} (left) Phase velocity and (right) group velocity 
from the 2-D CK solver versus wavenumbers, when $\Delta y=\Delta x$ and 
$c \Delta t = \Delta x$.}
\end{center}
\end{figure}

The maps of the 2-D phase and group velocities are plotted in Fig. \ref{fig:Vckc2d}, 
using the time step at the Courant limit $c\Delta t = \Delta x$, 
with $\Delta y=\Delta x$.
In this case, the phase and group velocities are exact along the main axes. 
At any angle from the axes, both the phase and group velocities are 
below the physical value, down to $c/\sqrt{2}$ and to $0$ for the phase and group velocities respectively,
at the Nyquist wavelength along the diagonal.

In 3-D, the phase and group velocities are exact along the main axes of the grid 
when using the time step at the Courant limit $c\Delta t = \Delta x$, 
with $\Delta z=\Delta y=\Delta x$. At other angles, both the phase and group velocities are 
below the physical value.

Hence, assuming cubic cells ($\Delta x=\Delta y=\Delta z$), the coefficients
given in \cite{KarkICAP06} ($\alpha=7/12$, $\beta=1/12$ and $\xi=1/48$)
allow for the Courant condition to be at $\Delta t=\Delta x$, which
equates to having no numerical dispersion along the principal axes.
The algorithm reduces to the FDTD algorithm with $\alpha=1$ and $\beta=\xi=0$.
Prescriptions for the coefficients are provided by Cowan, \emph{et al.}
in 3-D in \cite{CowanPRSTAB13} and by Pukhov in 2-D in
\cite{PukhovJPP99}, that enable no numerical dispersion along the 
direction of the smallest cell size when using non-cubic cells. 
An alternative NSFDTD implementation that enables superluminous waves along 
one axis is also given by Lehe {\it et al.} in \cite{LehePRSTAB13}. 

Enabling low or no numerical dispersion at all angles necessitates the use of higher-order 
finite-differences, which is more efficiently handled with Fourier-based spectral 
solvers.

\subsubsection{Pseudo Spectral Analytical Time Domain (PSATD)}

High-order approximations to the first spatial derivative can be obtained by 
using a larger stencil that involves more points on the grid. At the limit of 
infinite order, the approximate derivative is considered exact. It can 
be shown that at the limit of infinite order approximation of the spatial 
derivatives and infinitely small time steps, the algorithm is exact for 
all wavelengths and directions. 

Even for high finite order and small, 
but finite, sub time steps, such an algorithm would be too costly and 
impractical. 
Solving instead Maxwell's equation in Fourier space enables the evaluation of the 
derivative at any order, and analytical integration in time, thanks to the 
linearity of the equations.

Maxwell's equations in Fourier space are given by 
\begin{subequations}
\begin{eqnarray}
\frac{\partial\fe}{\partial t} & = & i\fk\times\fb-\fj\\
\frac{\partial\fb}{\partial t} & = & -i\fk\times\fe\\
{}[i\fk\cdot\fe & = & \tilde{\rho}]\\
{}[i\fk\cdot\fb & = & 0]
\end{eqnarray}
\end{subequations}
where $\tilde{a}$ is the Fourier Transform of the quantity $a$.
As with the real space formulation, provided that the continuity equation
$\partial\tilde{\rho}/\partial t+i\fk\cdot\fj=0$ is satisfied, then
the last two equations will automatically be satisfied at any time
if satisfied initially and do not need to be explicitly integrated.

Decomposing the electric field and current between longitudinal and
transverse components $\fe=\fe_{L}+\fe_{T}=\fkhat(\fkhat\cdot\fe)-\fkhat\times(\fkhat\times\fe)$
and $\fj=\fj_{L}+\fj_{T}=\fkhat(\fkhat\cdot\fj)-\fkhat\times(\fkhat\times\fj)$
gives
\begin{subequations}
\begin{eqnarray}
\frac{\partial\fe_{T}}{\partial t} & = & i\fk\times\fb-\mathbf{\tilde{J}_{T}}\\
\frac{\partial\fe_{L}}{\partial t} & = & -\mathbf{\tilde{J}_{L}}\\
\frac{\partial\fb}{\partial t} & = & -i\fk\times\fe
\end{eqnarray}
\end{subequations}
with $\fkhat=\fk/k$.

If the sources are assumed to be constant over a time interval $\Delta t$,
the system of equations is solvable analytically and is given by (see
\cite{HaberICNSP73} for the original formulation and \cite{VayJCP13}
for a more detailed derivation):
%
\begin{subequations}
\label{Eq:PSATD}
\begin{eqnarray}
\fe_{T}^{n+1} & = & C\fe_{T}^{n}+iS\fkhat\times\fb^{n}-\frac{S}{k}\fj_{T}^{n+1/2}\label{Eq:PSATD_transverse_1}\\
\fe_{L}^{n+1} & = & \fe_{L}^{n}-\Delta t\fj_{L}^{n+1/2}\\
\fb^{n+1} & = & C\fb^{n}-iS\fkhat\times\fe^{n}\nonumber\\
&+&i\frac{1-C}{k}\fkhat\times\fj^{n+1/2}\label{Eq:PSATD_transverse_2}
\end{eqnarray}
\end{subequations}
with $C=\cos\left(k\Delta t\right)$ and $S=\sin\left(k\Delta t\right)$.

Combining the transverse and longitudinal components gives 
\begin{subequations}
\begin{eqnarray}
\fe^{n+1} & = & C\fe^{n}+iS\fkhat\times\fb^{n}-\frac{S}{k}\fj^{n+1/2}\nonumber\\
 & + &(1-C)\fkhat(\fkhat\cdot\fe^{n})\nonumber \\
 & + & \fkhat(\fkhat\cdot\fj^{n+1/2})\left(\frac{S}{k}-\Delta t\right),\label{Eq_PSATD_1}\\
\fb^{n+1} & = & C\fb^{n}-iS\fkhat\times\fe^{n}\\
&+&i\frac{1-C}{k}\fkhat\times\fj^{n+1/2}.\label{Eq_PSATD_2}
\end{eqnarray}
\end{subequations}

For fields generated by the source terms without the self-consistent
dynamics of the charged particles, this algorithm is free of numerical
dispersion and is not subject to a Courant condition. Furthermore,
this solution is exact for any time step size, subject to the initial assumption
that the current source is constant over that time step, which turns out to be 
a standard assumption in the Particle-In-Cell method. 


The PSATD formulation that was just given applies to the
field components located at the nodes of the grid. As noted in \cite{Ohmurapiers2010},
they can also be easily recast on a staggered Yee grid by multiplication
of the field components by the appropriate phase factors to shift
them from the collocated to the staggered locations. The choice between
a collocated and a staggered formulation is application-dependent.

\subsubsubsection{Finite-order PSATD-p method}

Spectral solvers used to be very popular in the years 1970s to early 1990s, before being replaced by finite-difference methods with the advent of parallel supercomputers that favored local methods. However, it was shown recently that standard domain decomposition with Fast Fourier Transforms that are local to each subdomain could be used effectively with PIC spectral methods \cite{VayJCP13}, at the cost of truncation errors in the guard cells that could be neglected under some conditions. 
Furthermore, using very high - but finite - order enables the use of pseudo-spectral solver with ultrahigh accuracy and 
a compact support that permits domain decomposition and scaling to a very large number of subdomains, for 
parallel simulations on multiple computer nodes (see Section \ref{Sec:HPC}).
A detailed analysis of the effectiveness of the method with exact evaluation of the magnitude of the effect of the truncation error is given in \cite{Vincenti2016a} for stencils of arbitrary order (up-to the infinite ``spectral'' order).

The PSATD algorithm is generalized to an arbitrary order $p$ by simply substituting the exact spatial derivatives 
in Fourier space by the representation of the $p$-order finite-difference of the operator in Fourier space. I.e., 
the first derivative of the quantity $G$ along the axis $x$ in Fourier space is approximated by 
\begin{align}
k_x \tilde{G} = k^p_x \tilde{G} \;\;\; [+\mathcal{O}(\Delta x^p)],
\end{align}
with (for a staggered grid)
\begin{align}
k^p_x = \sum_{j=1}^{p/2}C^p_{j}\frac{\sin\left[\frac{(2j-1)k_{x}\Delta x}{2}\right]}{\Delta x/2},
\end{align}
where $C^p_j$ is the $j^{th}$ Fornberg coefficients (\cite{Fornberg1990}) at order $p$.
Similar approximations $k^p_y$ and $k^p_z$ are performed along the other axes.

\begin{figure}[ht]
\begin{center}
\includegraphics[trim={0.cm 7.cm 6cm 0.cm},clip,scale=0.5]{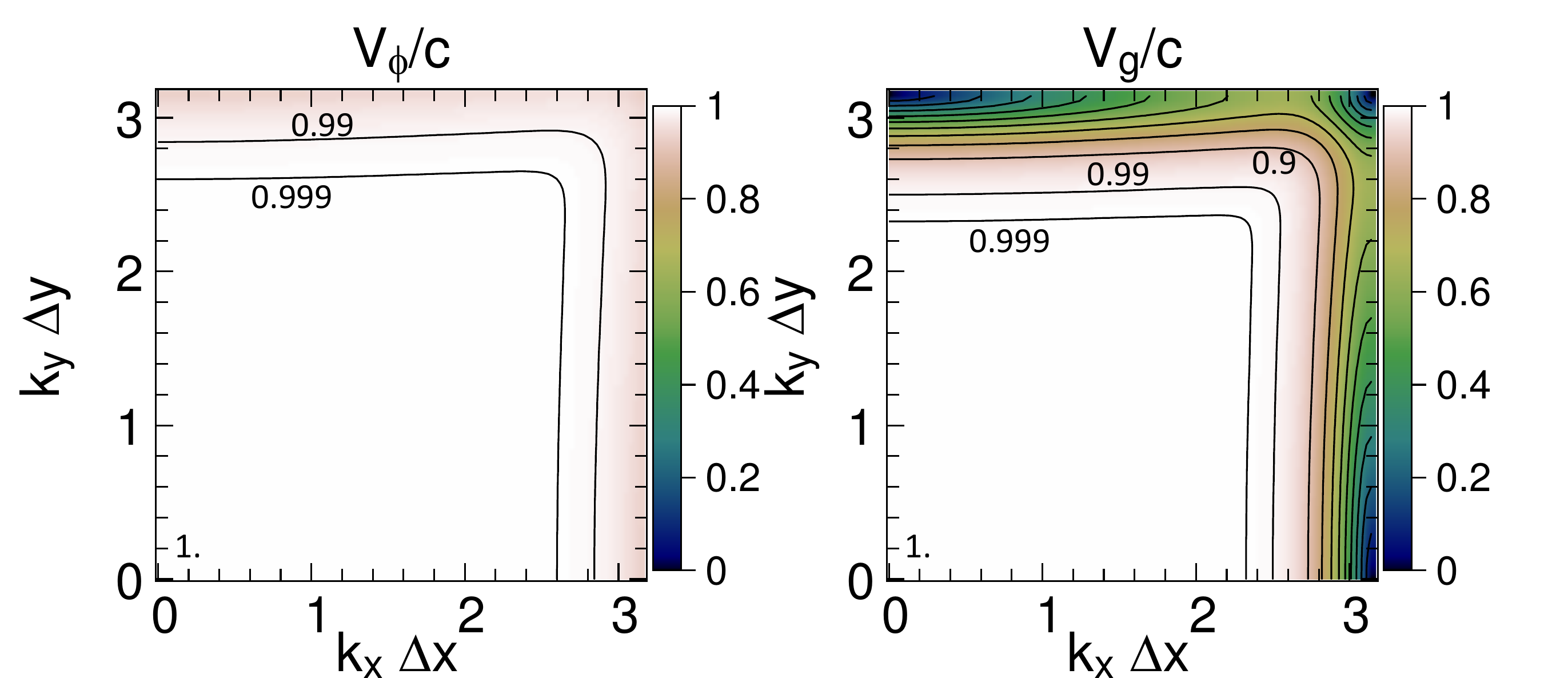}
\caption{\label{fig:Vpsatd_p64_2d} (left) Phase velocity and (right) group velocity 
from the 2-D PSATD solver at order $p=64$ versus wavenumbers, when $\Delta y=\Delta x$.}
\end{center}
\end{figure}

The phase and group velocities for the 2-D PSATD solver at order $p=64$, versus wavenumbers, 
are given in Fig. \ref{fig:Vpsatd_p64_2d} (assuming discretization on a staggered grid). 
In that case, the phase and group velocities have very low numerical dispersion at all angles and 
a very wide fraction of the spectrum. Note that in this case, since the algorithm is based on 
an analytic integration in time, there is no CFL condition, and the phase and group velocities in 
vacuum are thus independent of the time step value.

\subsection{Current deposition}

The current densities are deposited on the computational grid from
the particle position and velocities, employing splines of various
orders \cite{AbeJCP86}.
\begin{subequations}
\begin{eqnarray}
\mathbf{J} & = & \frac{1}{\Delta x \Delta y \Delta z}\sum_i q_i\mathbf{v_i}S_x^{(n)}S_y^{(n)}S_z^{(n)},
\end{eqnarray}
\end{subequations}
where $S^{(n)}$ is a spline of order $n$ that is given by the $n^{th}$ convolution of 
the rectangular function $\prod$ by itself. Hence, the Fourier transform of 
the spline of order $n$ is given by
\begin{align}
\tilde{S}_x^{(n)}(k) = \left[\frac{sin\left(k_x\Delta x/2\right)}{k_x\Delta x/2}\right]^{n+1}.
\end{align}

\begin{figure}[ht]
\begin{center}
\includegraphics[trim={0.cm 0.cm 0cm 0.cm},clip,scale=0.5]{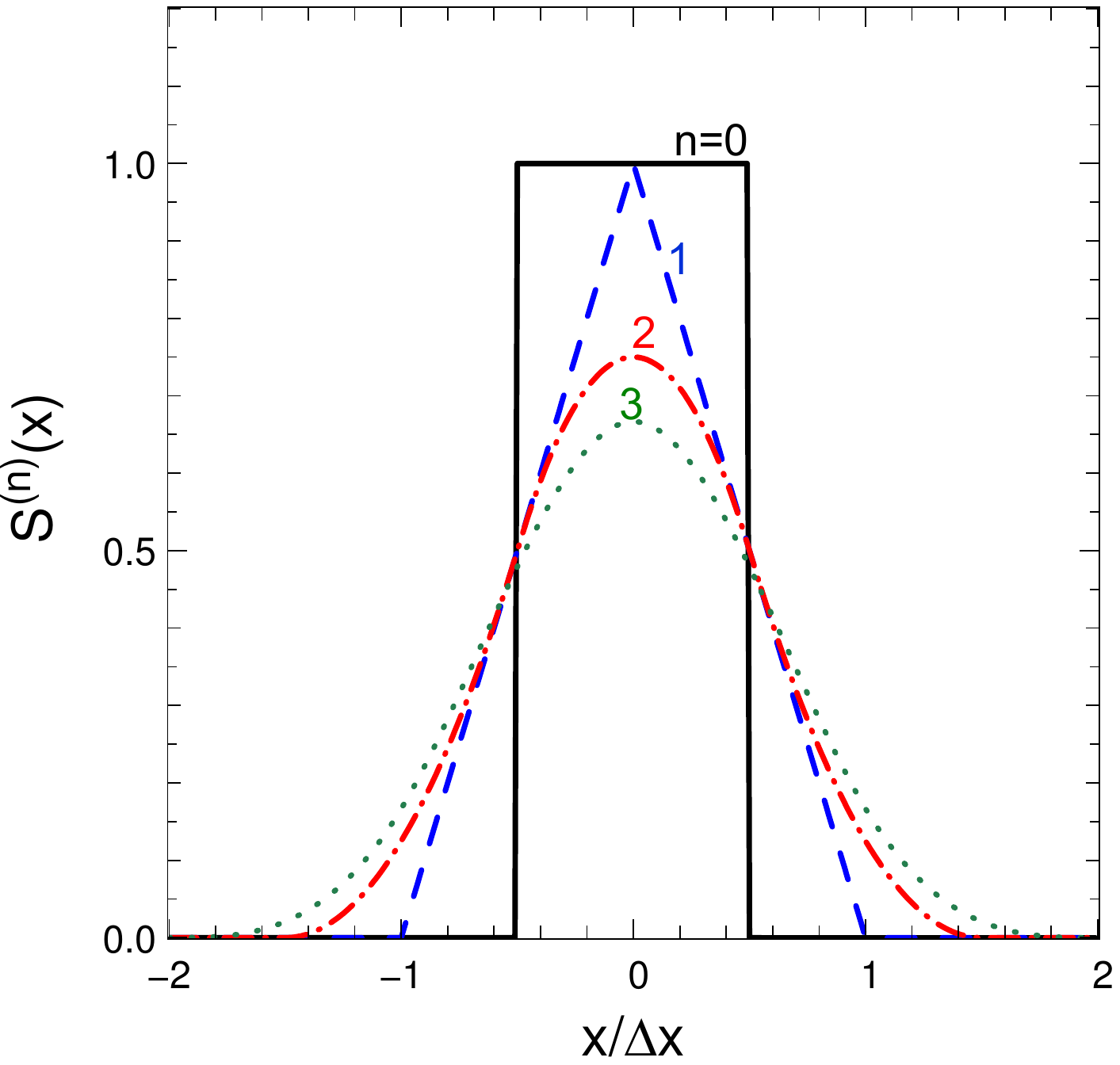}
\caption{\label{fig:splines} Splines $S^{(n)}$ for current deposition and force gathering for 
orders $0$ to $3$.}
\end{center}
\end{figure}

The splines for orders $n={0,1,2,3}$ are given by
\begin{subequations}
\begin{eqnarray}
S_x^{(0)} = & 1 & \;\;\; \text{for} \;\; 0 \leq |x| \leq \Delta x/2, \\
S_x^{(1)} = & \frac{\Delta x-|x|}{\Delta x} & \;\;\; \text{for} \;\; 0 \leq |x| \leq \Delta x, \\
S_x^{(2)} = & \frac{\frac{3}{4}\Delta x^2-x^2}{\Delta x^2} &\;\;\; \text{for} \;\; 0 \leq |x| \leq \Delta x/2,\\
                = & \frac{(\frac{3}{2}\Delta x-|x|)^2}{2\Delta x^2} &\;\;\; \text{for} \;\; \Delta x/2 \leq |x| \leq 3\Delta x/2, \\ 
S_x^{(3)} = & \frac{2}{3}-\frac{x^2 (1-|x|/2)}{\Delta x^3} &\;\;\; \text{for} \;\; 0 \leq |x| \leq \Delta x, \\
                = & \frac{(2\Delta x-|x|)^3}{6\Delta x^3} &\;\;\; \text{for} \;\; \Delta x \leq |x| \leq 2\Delta x,
\end{eqnarray}
\end{subequations}
and are plotted in Fig.\ref{fig:splines}.

In most applications, it is essential to prevent the accumulation
of errors resulting from the violation of the discretized Gauss' Law.
This is accomplished by providing a method for depositing the current
from the particles to the grid that preserves the discretized Gauss'
Law, or by providing a mechanism for ``divergence cleaning'' \cite{BirdsallLangdon,LangdonCPC92,MarderJCP87,VayPOP98,Munzjcp2000}.
For the former, schemes that allow a deposition of the current that
is exact when combined with the Yee solver is given in \cite{VillasenorCPC92}
for linear splines and in \cite{Esirkepovcpc01} for splines of arbitrary order. 

Note that the NSFDTD formulations given above and in \cite{PukhovJPP99,VayJCP2011,CowanPRSTAB13,LehePRSTAB13} 
apply to the Maxwell-Faraday
equation, while the discretized Maxwell-Ampere equation uses the FDTD
formulation. Consequently, the charge conserving algorithms developed
for current deposition \cite{VillasenorCPC92,Esirkepovcpc01} apply
readily to those NSFDTD-based formulations.  

\subsection{Field gather}

\begin{figure}[ht]
\begin{center}
\includegraphics[trim={0.cm 0.cm 0cm 0.cm},clip,scale=0.7]{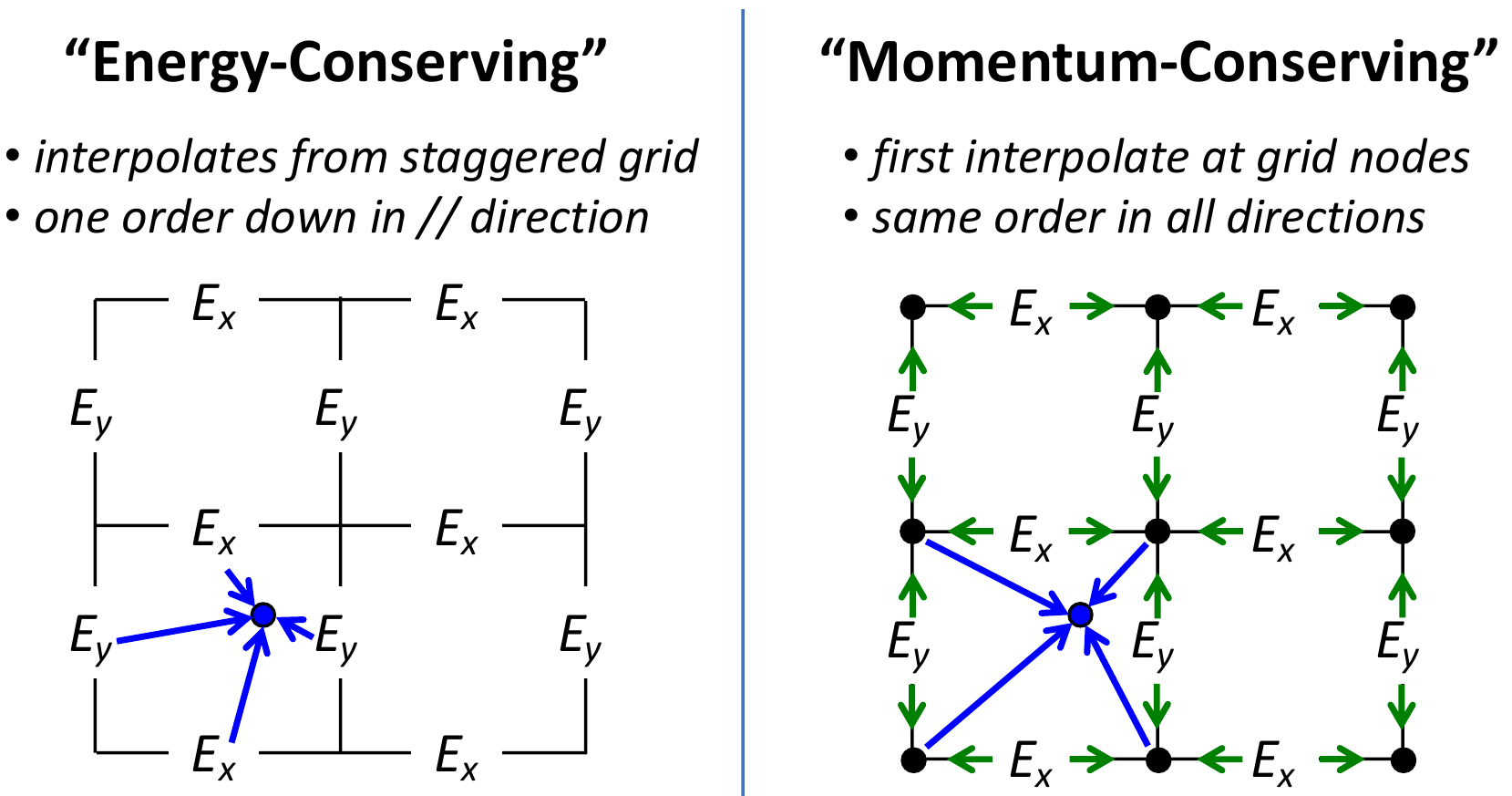}
\caption{\label{fig:gather_grids} (left) ``Energy-conserving'' (EC) and (right) ``momentum-conserving'' (MC)
schemes for gathering the electric fields from a 2-D staggered Yee mesh onto the particles, 
assuming linear splines ($S^{(1)}$). For the EC gather, the fields are interpolated directly 
from the staggered grid, using a spline of order $n-1$ (i.e. $S^{(0)}$) along $x$ for $E_x$ and along $y$ for $E_y$, 
and using a spline of order $n$ (i.e. $S^{(1)}$) along $y$ for $E_x$ and along $x$ for $E_y$.
For the MC gather, the fields are first interpolated (using linear interpolation) from the staggered positions 
onto the nodes of the grids, then to the particles positions using $S^{(1)}$ in all directions for every component. See 
text for interpolation of the magnetic field components.
}
\end{center}
\end{figure}

In general, the field is gathered from the mesh onto the macroparticles
using splines of the same order as for the current deposition $\mathbf{S}=\left(S_{x},S_{y},S_{z}\right)$.
Two variations are considered:
\begin{itemize}
\item ``energy conserving'': fields are interpolated from
the staggered Yee grid to the macroparticles using 
$\left(S^{(n-1)}_x,S^{(n)}_y,S^{(n)}_z\right)$ for $E_{x}$, 
$\left(S^{(n)}_x,S^{(n-1)}_y,S^{(n)}_z\right)$ for $E_{y}$,
$\left(S^{(n)}_x,S^{(n)}_y,S^{(n-1)}_z\right)$ for $E_{z}$, 
$\left(S^{(n)}_x,S^{(n-1)}_y,S^{(n-1)}_z\right)$ for $B_{x}$, 
$\left(S^{(n-1)}_x,S^{(n)}_y,S^{(n-1)}_z\right)$ for $B_{y}$ and 
$\left(S^{(n-1)}_x,S^{(n-1)}_y,S^{(n)}_z\right)$ for $B_{z}$,
\item ``momentum conserving'': fields are interpolated from the grid nodes
to the macroparticles using $\mathbf{S}=\left(S^{(n)}_x,S^{(n)}_y,S^{(n)}_z\right))$
for all field components (since the fields are known at staggered positions,
they are first interpolated to the nodes, usually using linear interpolation).
\end{itemize}
Diagrams illustrating the two schemes in the standard case of linear splines ($S^{(1)}$) 
are given in Fig.\ref{fig:gather_grids}.

As shown in \cite{BirdsallLangdon,HockneyEastwoodBook,LewisJCP1972},
the energy and momentum conserving schemes conserve energy or momentum 
respectively at the limit of infinitesimal time steps and generally
offer better conservation of the respective quantities for a finite
time step. Neither method is intrinsically superior to the other, and 
it can be useful to implement both methods and test whether one 
converges faster than the other for a given class of problems.

\subsection{Filtering \label{Sec:filtering}}

It is common practice to apply digital filtering to the
current density in Particle-In-Cell simulations as a complement or
an alternative to using higher order splines \cite{BirdsallLangdon}.
As seen above in the section on the field solvers, the group velocity 
often vanishes at the Nyquist wavelength, and the Leapfrog Maxwell 
integrator is not stable at this wavelength. It is thus prudent to completely 
remove the Nyquist wavelength from the source term. Since the 
group velocity of the neighboring wavelengths are usually very inaccurate, 
it may also be beneficial to damp them.

\begin{figure}[ht]
\begin{center}
\includegraphics[trim={0.cm 0.cm 0cm 0.cm},clip,scale=0.7]{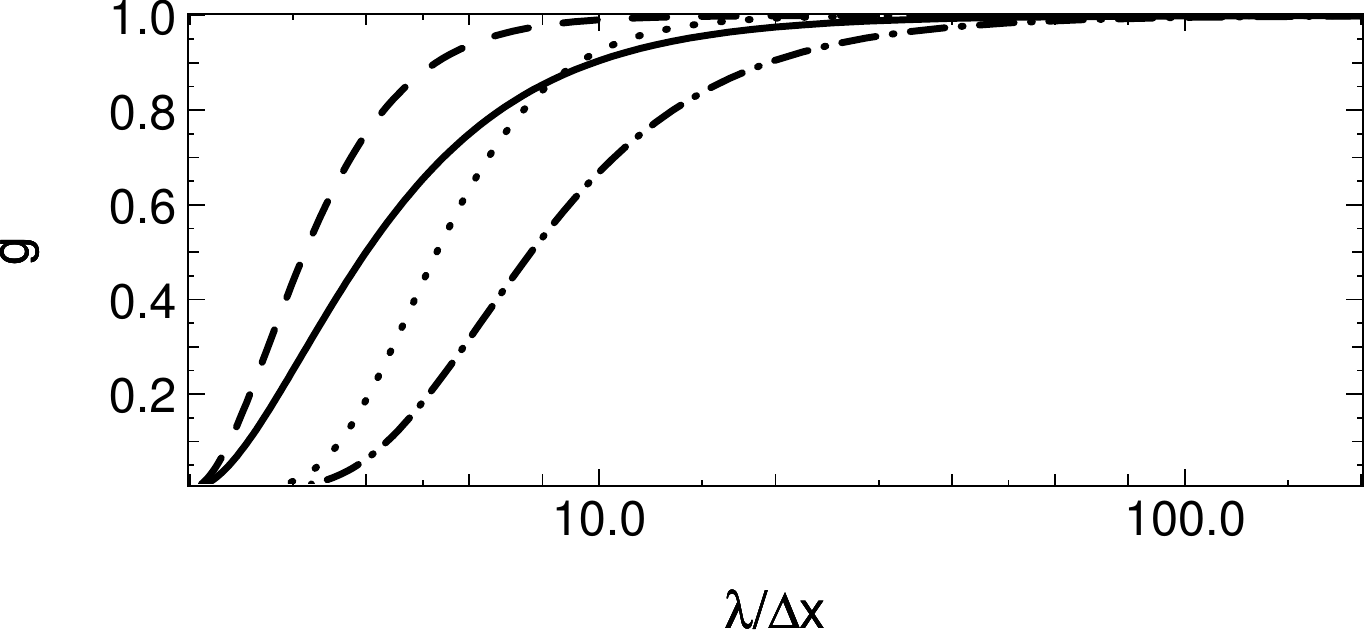}
\caption{\label{fig:filter} Gain of (solid) one pass of bilinear filter without 
compensation, (dash) one pass of bilinear filter with compensation, 
(dot-dash) four passes of bilinear filter without  compensation,
(dot) four passes of bilinear filter with  compensation.
}
\end{center}
\end{figure}

A commonly used filter in PIC simulations is the three points filter
\begin{align}
\phi_{j}^{f}=\alpha\phi_{j}+\left(1-\alpha\right)\left(\phi_{j-1}+\phi_{j+1}\right)/2,
\end{align}
where $\phi^{f}$ is the filtered quantity. This filter is called
a ``bilinear'' filter when $\alpha=0.5$. Assuming $\phi=e^{\i kx}$ and
$\phi^{f}=g\left(\alpha,k\right)e^{\i kx}$, the filter gain $g$ is
given as a function of the filtering coefficient $\alpha$ and
the wavenumber $k$ by 
\begin{align}
g\left(\alpha,k\right)=\alpha+\left(1-\alpha\right)\cos\left(k\Delta x\right)\approx1-\left(1-\alpha\right)\frac{\left(k\Delta x\right)^{2}}{2}+O\left(k^{4}\right).
\end{align}
The total attenuation $G$ for $n$ successive applications of filters
of coefficients $\alpha_{1}$...$\alpha_{n}$ is given by $G=\prod_{i=1}^{n}g\left(\alpha_{i},k\right)\approx1-\left(n-\sum_{i=1}^{n}\alpha_{i}\right)\frac{\left(k\Delta x\right)^{2}}{2}+O\left(k^{4}\right)$.
A sharper cutoff in $k$ space is provided by using $\alpha_{n}=n-\sum_{i=1}^{n-1}\alpha_{i}$,
so that $G\approx1+O\left(k^{4}\right)$. Such step is called a ``compensation''
step \cite{BirdsallLangdon}. For the bilinear filter ($\alpha=1/2$),
the compensation factor is $\alpha_{c}=2-1/2=3/2$. For a succession
of $n$ applications of the bilinear factor, it is $\alpha_{c}=n/2+1$. 

Examples of gains of commonly used filters are shown in Fig.\ref{fig:filter}.

\subsection{Energy and momentum conservation}

\begin{figure}[ht]
\begin{center}
\includegraphics[trim={0.cm 0.cm 0cm 0.cm},clip,scale=0.8]{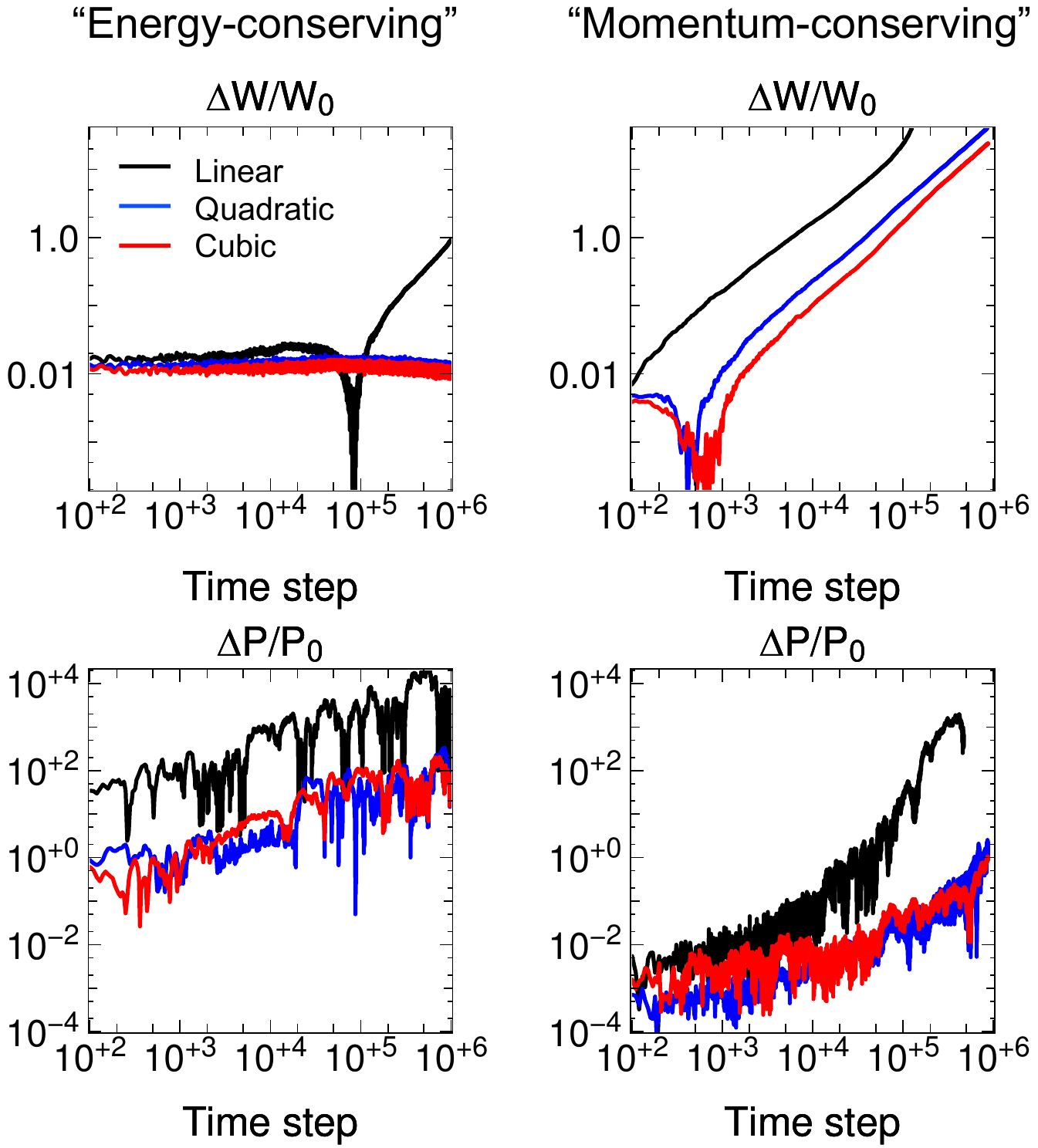}
\caption{\label{fig:Whist}
Evolutions of the total relative energy (top) and momentum (bottom) from 2-D 
electromagnetic PIC simulations with the Yee Maxwell solver, using either 
the ``energy-conserving'' (left) or the ``momentum-conserving'' (right) 
gather, with linear (black), quadratic (blue) or cubic (red) splines.
}
\end{center}
\end{figure}

As mentioned above, and discussed in detail in \cite{BirdsallLangdon,HockneyEastwoodBook}, there are schemes 
that can preserve energy or momentum, but only at the infinitesimal limit. When using 
finite time steps, as required in practice, neither is strictly preserved. It is thus 
important to have in mind these limitations and explore what set of parameters 
work best for a given problem.

To illustrate the energy and momentum conservation properties of the standard 
PIC using the Yee algorithms, histories of the global energy and momentum are 
given in Fig. \ref{fig:Whist} from simulations of a uniform warm plasma at rest, using either 
the ``energy-conserving'' or the ``momentum-conserving'' algorithm with 
deposition and gathering splines of orders 1 (linear), 2 (quadratic) or 3 (cubic).

As expected, the simulations using the ``energy-conserving'' gather do 
better at preserving the energy while the ones using the ``momentum-conserving'' 
gather preserved the momentum better. As explained above, the linear growth of energy is due 
to numerical ``stochastic heating'' from random errors from the fields acting on the 
particle motion. Using higher order splines reduces the magnitude of these random 
errors and thus the growth rate \cite{AbeJCP86,HockneyEastwoodBook,BirdsallLangdon}.

\section{Application to the modeling of plasma-based accelerators}

\subsection{Moving window and optimal Lorentz boosted frame}
The simulations of plasma accelerators from first principles are extremely computationally intensive, due to the need to resolve the evolution of a driver (laser or particle beam) and an accelerated particle beam into a plasma structure that is orders of magnitude longer and wider than the accelerated beam. As is customary in the modeling of particle beam dynamics in standard particle accelerators, a moving window is commonly used to follow the driver, the wake and the accelerated beam. This results in huge savings, by avoiding the meshing of the entire plasma that is orders of magnitude longer than the other length scales of interest. 

\begin{figure}[ht]
\begin{center}
\includegraphics[trim={1.cm 11cm 1cm 3cm},clip,scale=0.6]{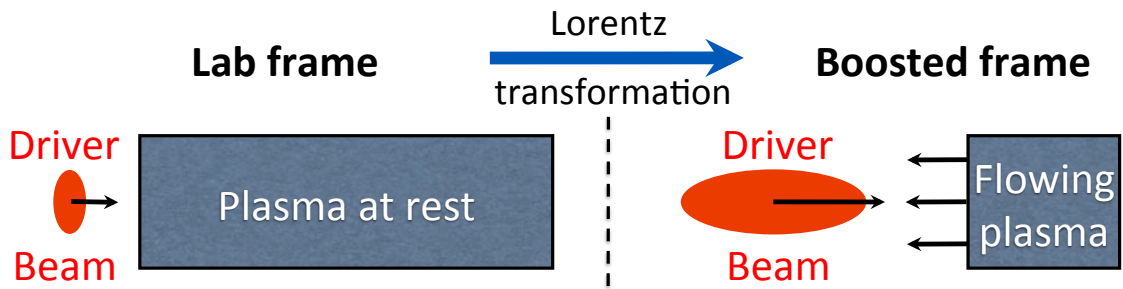}
\caption{\label{fig:BF} A first principle simulation of a short driver beam (laser or charged particles) propagating through a plasma that is orders of magnitude longer necessitates a very large number of time steps. Recasting the simulation in a frame of reference that is moving close to the speed of light in the direction of the driver beam leads to simulating a driver beam that appears longer propagating through a plasma that appears shorter than in the laboratory. Thus, this relativistic transformation of space and time reduces the disparity of scales, and thereby the number of time steps to complete the simulation, by orders of magnitude.}
\end{center}
\end{figure}

Even using a moving window, however, a full PIC simulation of a plasma accelerator can be extraordinarily demanding computationally, as many time steps are needed to resolve the crossing of the short driver beam with the plasma column. As it turns out, choosing an optimal frame of reference that travels close to the speed of light in the direction of the laser or particle beam (as opposed to the usual choice of the laboratory frame) enables speedups by orders of magnitude \cite{VayPRL07,VayPOP2011}. This is a result of the properties of Lorentz contraction and dilation of space and time. In the frame of the laboratory, a very short driver (laser or particle) beam propagates through a much longer plasma column, necessitating millions to tens of millions of time steps for parameters in the range of the BELLA or FACET-II experiments. As sketched in Fig. \ref{fig:BF}, in a frame moving with the driver beam in the plasma at velocity $v=\beta c$ (where $c$ is the speed of light in vacuum), the beam length is now elongated by $\approx(1+\beta)\gamma$ while the plasma contracts by $\gamma$ (where $\gamma=1/\sqrt{1-\beta^2}$ is the relativistic factor associated with the frame velocity). The number of time steps that is needed to simulate a ``longer'' beam through a ``shorter'' plasma is now reduced by up to $\approx(1+\beta) \gamma^2$ (a detailed derivation of the speedup is given in \cite{VayPOP2011}). For simulations of multi-GeV stages in the linear regime, the group velocity of the laser 
can reach $\gamma>100$, leading to speedups $(1+\beta)\gamma^2>20000$.

The modeling of a plasma acceleration stage in a boosted frame 
involves the fully electromagnetic modeling of a plasma propagating at near the speed of light, for which Numerical Cerenkov 
\cite{BorisJCP73,HaberICNSP73} is a potential issue, as explained in more details below.
In addition, for a frame of reference moving in the direction of the accelerated beam (or equivalently the wake of the laser), 
waves emitted by the plasma in the forward direction expand 
while the ones emitted in the backward direction contract, following the properties of the Lorentz transformation. 
If one had to resolve both forward and backward propagating 
waves emitted from the plasma, there would be no gain in selecting a frame different from the laboratory frame. However, 
the physics of interest for a laser wakefield is the laser driving the wake, the wake, and the accelerated beam. 
Backscatter is weak in the short-pulse regime, and does not 
interact as strongly with the beam as do the forward propagating waves 
which stay in phase for a long period. It is thus often assumed that the backward propagating waves 
can be neglected in the modeling of plasma accelerator stages. The accuracy  of this assumption has been demonstrated by 
comparison between explicit codes that include both forward and backward waves with envelope or quasistatic 
codes that neglect backward waves \cite{GeddesJP08,GeddesPAC09,CowanAAC08}.

\subsection{Numerical Cherenkov Instability and alternate formulation in a Galilean frame}

\begin{figure}[ht]
\begin{center}
\includegraphics[trim={0.cm 0cm 0cm 0cm},clip,scale=0.6]{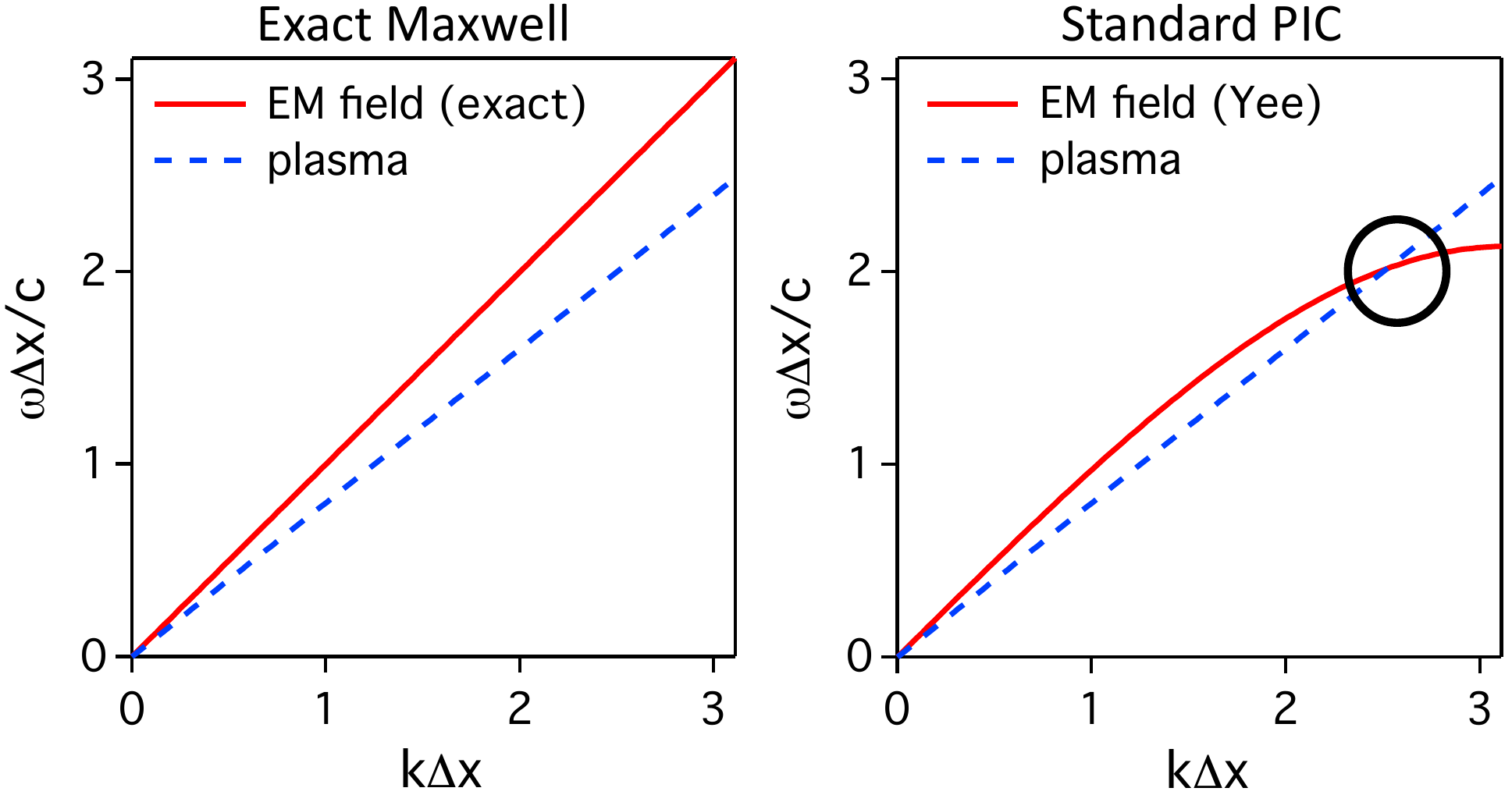}
\includegraphics[trim={0.cm 0cm 0cm 0cm},clip,scale=0.5]{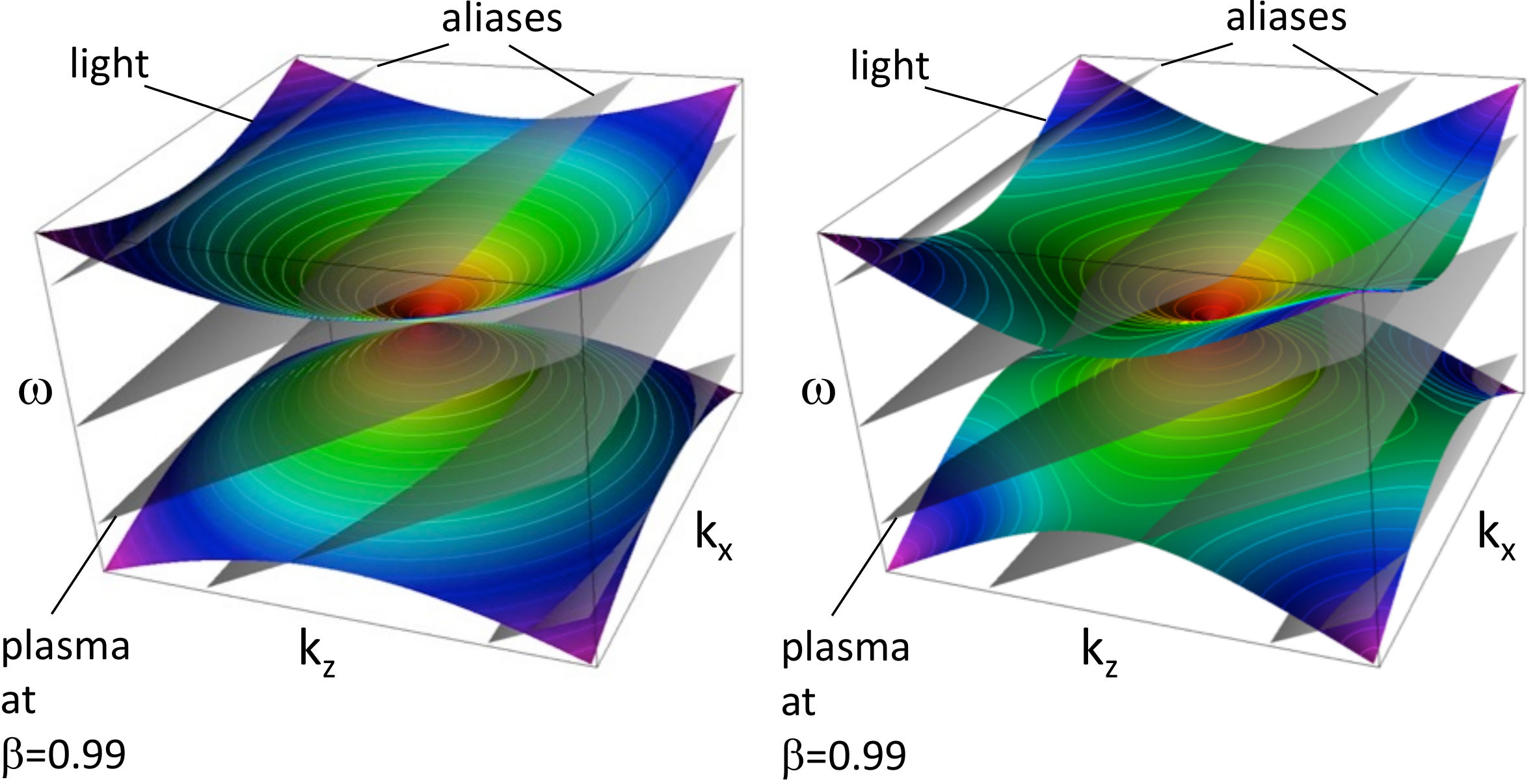}
\caption{\label{fig:NCI} 
(top-left) Exact dispersion relations in 1-D of light in vacuum (solid red line) and 
a relativistically drifting plasma (or particle beam). There is no crossing except at 
the origin and the system is stable. (top-right) Dispersion relations in 1-D for a 
PIC simulation with the Yee Maxwell solver with $c \Delta t = \Delta x /\sqrt{2}$. 
The distortion from numerical dispersion induces a crossing (black circle) that leads to 
an instability. Aliases from the space and time discretization lead to additional crossing 
and unstable modes, whether using an exact Maxwell solver (i.e. PSATD) with accurate 
light cones (bottom-left) or a solver with numerical dispersion such as the Yee 
solver (bottom-right).
}
\end{center}
\end{figure}

The ``Numerical Cherenkov Instability'' (NCI) \cite{godfreyjcp74}
is the most serious numerical instability affecting multidimensional
PIC simulations of relativistic particle beams and streaming plasmas
\cite{MartinsCPC10,VayAAC2010,VayJCP2011,Spitkovsky:Icnsp2011,GodfreyJCP2013,XuJCP2013}. 
It arises from coupling between possibly numerically distorted electromagnetic modes and spurious
beam modes, the latter due to the mismatch between the Lagrangian
treatment of particles and the Eulerian treatment of fields \cite{Godfreyjcp75} (See Fig. \ref{fig:NCI}).
In recent papers, the electromagnetic dispersion
relations for the numerical Cherenkov instability were derived and solved for both FDTD \cite{GodfreyJCP2013,GodfreyJCP2014_FDTD}
and PSATD \cite{GodfreyJCP2014_PSATD,GodfreyIEEE2014} algorithms. 

Several solutions have been proposed to mitigate the NCI \cite{GodfreyJCP2014,GodfreyIEEE2014,GodfreyJCP2014_PSATD,GodfreyCPC2015,YuCPC2015,YuCPC2015-Circ}. Although
these solutions reduce efficiently the growth rate of the numerical instability,
they typically introduce either strong smoothing of the currents and
fields, or arbitrary numerical corrections, which are
tuned specifically against the NCI and go beyond the
natural discretization of the underlying physical equation. Therefore,
it is sometimes unclear to what extent these added corrections could impact the
physics at stake for a given resolution.

For instance, NCI-specific corrections include periodically smoothing 
the electromagnetic field components \cite{MartinsCPC10}, 
using a special time step \cite{VayAAC2010,VayJCP2011} or
applying a wide-band smoothing of the current components \cite{
  VayAAC2010,VayJCP2011,VayPOPL2011}. Another set of mitigation methods
involves scaling the deposited
currents by a carefully-designed wavenumber-dependent factor
\cite{GodfreyJCP2014_FDTD,GodfreyIEEE2014} or slightly modifying the
ratio of electric and magnetic fields ($E/B$) before gathering their
value onto the macroparticles 
\cite{GodfreyJCP2014_PSATD,GodfreyCPC2015}. 
Yet another set of NCI-specific corrections
\cite{YuCPC2015,YuCPC2015-Circ} consists 
in combining a small timestep $\Delta t$, a sharp low-pass spatial filter,
and a spectral or high-order scheme that is tuned so as to
create a small, artificial ``bump'' in the dispersion relation
\cite{YuCPC2015}. While most mitigation methods have only been applied
to Cartesian geometry, this last
set of methods (\cite{YuCPC2015,YuCPC2015-Circ}) 
has the remarkable property that it can be applied
\cite{YuCPC2015-Circ} to both Cartesian geometry and
quasi-cylindrical geometry (i.e. cylindrical geometry with
azimuthal Fourier decomposition \cite{LifschitzJCP2009,DavidsonJCP2015,Lehe2016}). However,
the use of a small timestep proportionally slows down the progress of
the simulation, and the artificial ``bump'' is again an arbitrary correction
that departs from the underlying physics.

A new scheme was recently proposed, in \cite{Kirchen2016a,Lehe2016a}, which 
completely eliminates the NCI for a plasma drifting at a uniform relativistic velocity 
-- with no arbitrary correction -- by simply integrating
the PIC equations in \emph{Galilean coordinates} (also known as
\emph{comoving coordinates}). More precisely, in the new 
method, the Maxwell equations \emph{in Galilean coordinates} are integrated
analytically, using only natural hypotheses, within the PSATD
framework (Pseudo-Spectral-Analytical-Time-Domain \cite{HaberICNSP73,VayJCP2013}). 

The idea of the proposed scheme is to perform a Galilean change of
coordinates, and to carry out the simulation in the new coordinates:
\begin{equation} 
\label{eq:change-var}
\vec{x}' = \vec{x} - \vgal t 
\end{equation}
where $\vec{x} = x\,\vec{u}_x + y\,\vec{u}_y + z\,\vec{u}_z$ and
$\vec{x}' = x'\,\vec{u}_x + y'\,\vec{u}_y + z'\,\vec{u}_z$ are the
position vectors in the standard and Galilean coordinates
respectively. The new equations and algorithm derived 
in the Galilean frame are given in Appendix \ref{Sec:Gal}.

As shown in \cite{Kirchen2016a,Lehe2016a}, 
the elimination of the NCI with the new Galilean integration is verified empirically via PIC simulations of uniform drifting plasmas and laser-driven plasma acceleration stages, and confirmed by a theoretical analysis of the instability.

When choosing $\vgal= \vec{v}_0$, where
$\vec{v}_0$ is the speed of the bulk of the relativistic
plasma, the plasma does not move with respect to the grid in the Galilean 
coordinates $\vec{x}'$ -- or, equivalently, in the standard
coordinates $\vec{x}$, the grid moves along with the plasma. The heuristic intuition behind this scheme
is that these coordinates should prevent the discrepancy between the Lagrangian and
Eulerian point of view, which gives rise to the NCI \cite{Godfreyjcp75}.

An important remark is that the Galilean change of
coordinates (\ref{eq:change-var}) is a simple translation. Thus, when used in
the context of Lorentz-boosted simulations, it does
of course preserve the relativistic dilatation of space and time which gives rise to the
characteristic computational speedup of the boosted-frame technique.

Another important remark is that the Galilean scheme is \emph{not}
equivalent to a moving window (and in fact the Galilean scheme can be
independently \emph{combined} with a moving window). Whereas in a
moving window, gridpoints are added and removed so as to effectively
translate the boundaries, in the Galilean scheme the gridpoints
\emph{themselves} are not only translated but in this case, the physical equations
are modified accordingly. Most importantly, the assumed time evolution of
the current $\vec{J}$ within one timestep is different in a standard PSATD scheme with moving
window and in a Galilean PSATD scheme \cite{Lehe2016a}.

\subsection{Examples}

\subsubsection{3-D example of LWFA and PWFA}

\begin{figure}[ht]
\begin{center}
\includegraphics[trim={0.cm 0.cm 0cm 0.cm},clip,scale=1.2]{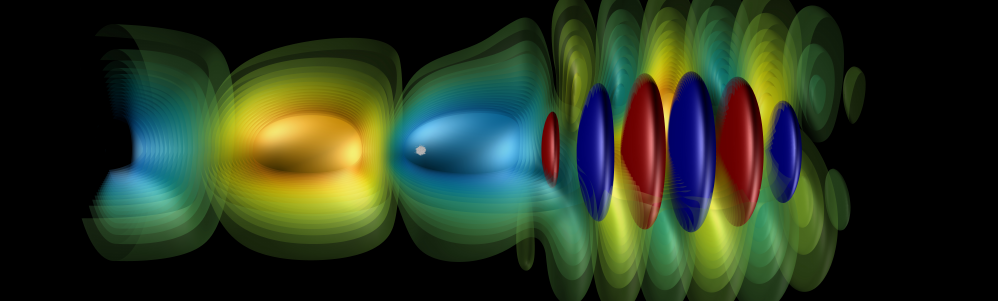}
\includegraphics[trim={0.cm 0.cm 0cm 0.cm},clip,scale=1.2]{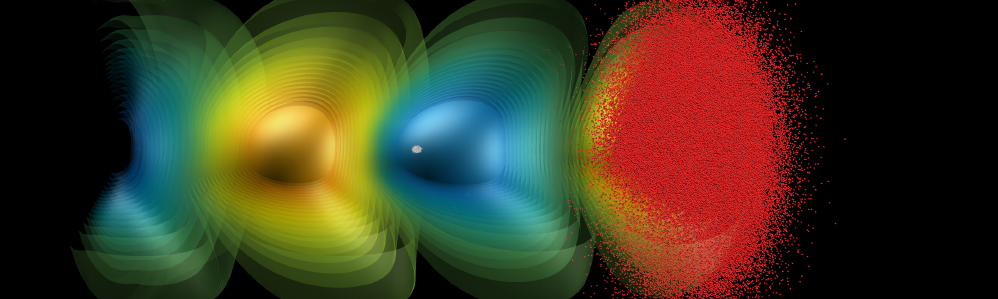}
\caption{\label{fig:LWFAPWFA} Renderings from 3-D simulations of (top) laser-driven, and 
(bottom) charged particle beam-driven, plasma accelerators. The laser driver 
(dark blue and red isosurfaces of the transverse electric field) and 
the particle beam driver (red macroparticles), are propagating from left to right.
The drivers disturb initially immobile electrons from a plasma column (not shown), 
through which they are propagating, generating plasma oscillations that results 
in a wake with longitudinal fields (light blue and yellow 
isosurfaces) of very high amplitude. An electron beam (white) placed at the right phase in the first accelerating 
bucket (light blue) is accelerated by the wake, gaining high energy in a short distance.
}
\end{center}
\end{figure}

The application of the standard electromagnetic PIC method is illustrated here with 
3-D simulations of a laser-driven and a charged-particle-beam-driven plasma accelerator.
Renderings are shown in Fig. \ref{fig:LWFAPWFA}, showing the driver beams and 
the wakes created by the driver propagating through the plasma column. 

These examples are from simulations using a moving window in the 
laboratory frame. They used the Yee Maxwell solver, ``energy-conserving'' gather, 
splines of order 3 and smoothing in the longitudinal direction (4 passes of 
bilinear filter + 1 pass of compensation).

\subsubsection{Diagnostics from boosted frame simulations}

\begin{figure}[ht]
\begin{center}
\includegraphics[trim={0.cm 0.cm 0cm 0.cm},clip,scale=0.57]{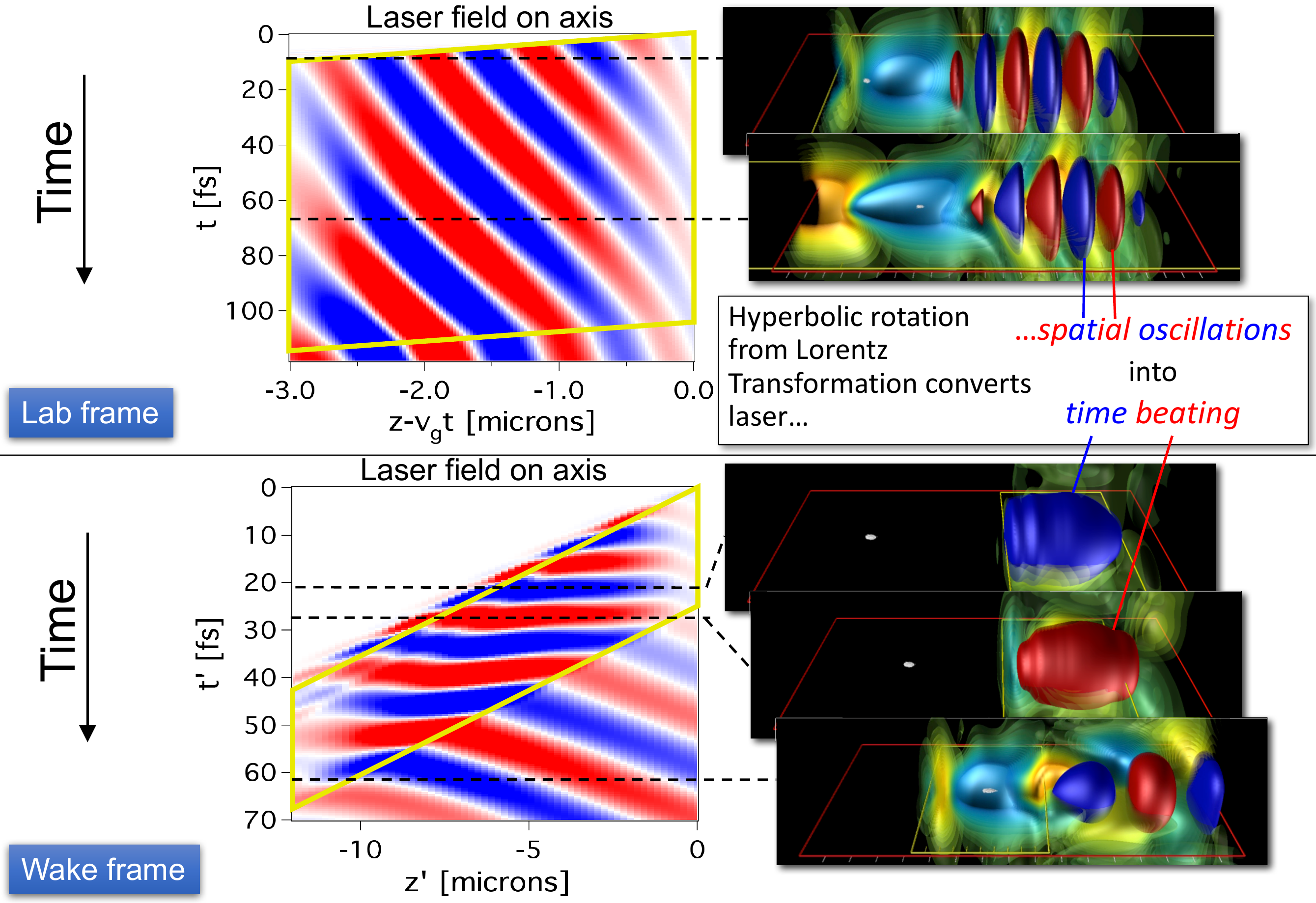}
\caption{\label{fig:BFLWFA}
(left) Space-time diagrams of the laser field amplitude on axis in the moving 
window following the laser in the plasma for (top) a simulation in the laboratory 
frame, and (bottom) for a simulation in a Lorentz boosted frame propagating 
at the velocity of the laser in the plasma. (right) Snapshots from the 
simulations at various times. The simulation box boundaries are delimited in the 
horizontal plane by the yellow lines in the snapshots. The plasma column boundaries 
are delimited by the red lines.
The hyperbolic rotation of quantities 
from the Lorentz Transformation between frames converts the 
laser spatial oscillations commonly observed in the laboratory frame 
to time beating in the boosted frame of the laser in the plasma (``wake frame'').
The usual spatial oscillations naturally occur again as the laser exits the plasma 
(right of the yellow rectangle in the bottom snapshot).
}
\end{center}
\end{figure}

As explained above, simulating in a Lorentz boosted frame can 
speedup the simulations by orders of magnitude when using extra 
precaution to control the numerical Cherenkov instability. 
In addition, input and output need special treatment to 
convert the input data from the laboratory frame to the boosted 
frame and the output data from the boosted frame to the laboratory.

In addition to analyzing the output data in the laboratory frame, it is 
instructive to analyze the data in the boosted frame, as the physics 
looks different and may lead to alternate insight for the physical 
phenomena as well as for numerical considerations.
It is especially true for the modeling of laser-driven plasma accelerators, 
where the group velocity of the laser in the plasma is usually low enough to be used 
as the Lorentz boosted frame. 

This is illustrated in Fig. \ref{fig:BFLWFA} that shows snapshots of the same 
3-D simulation of the previous section in the laboratory frame, as well 
as from a simulation of the same setup in a Lorentz boosted frame propagating 
at the group velocity of the laser in the plasma. The time histories of the 
laser field oscillations on axis are also shown for each simulation.
While spatial laser oscillations are clearly visible while the laser is propagating 
through the plasma in the laboratory frame simulation, the hyperbolic rotation 
of the Lorentz transformation converts them into pure time beating while 
the laser is propagating through the plasma in the boosted frame simulation. The usual spatial oscillations 
naturally occur again as the laser exits the plasma. One important consequence 
of this observation is that the relative content in short wavelength is reduced in 
boosted frame simulations, enabling more aggressive smoothing (if needed) 
\cite{VayPOPL2011} or potentially a coarser longitudinal resolution (relative to the 
laser wavelength in vacuum).

\subsubsection{Convergence}
\begin{figure}[ht]
\begin{center}
\includegraphics[trim={0.cm 0.cm 0cm 0.cm},clip,scale=0.25]{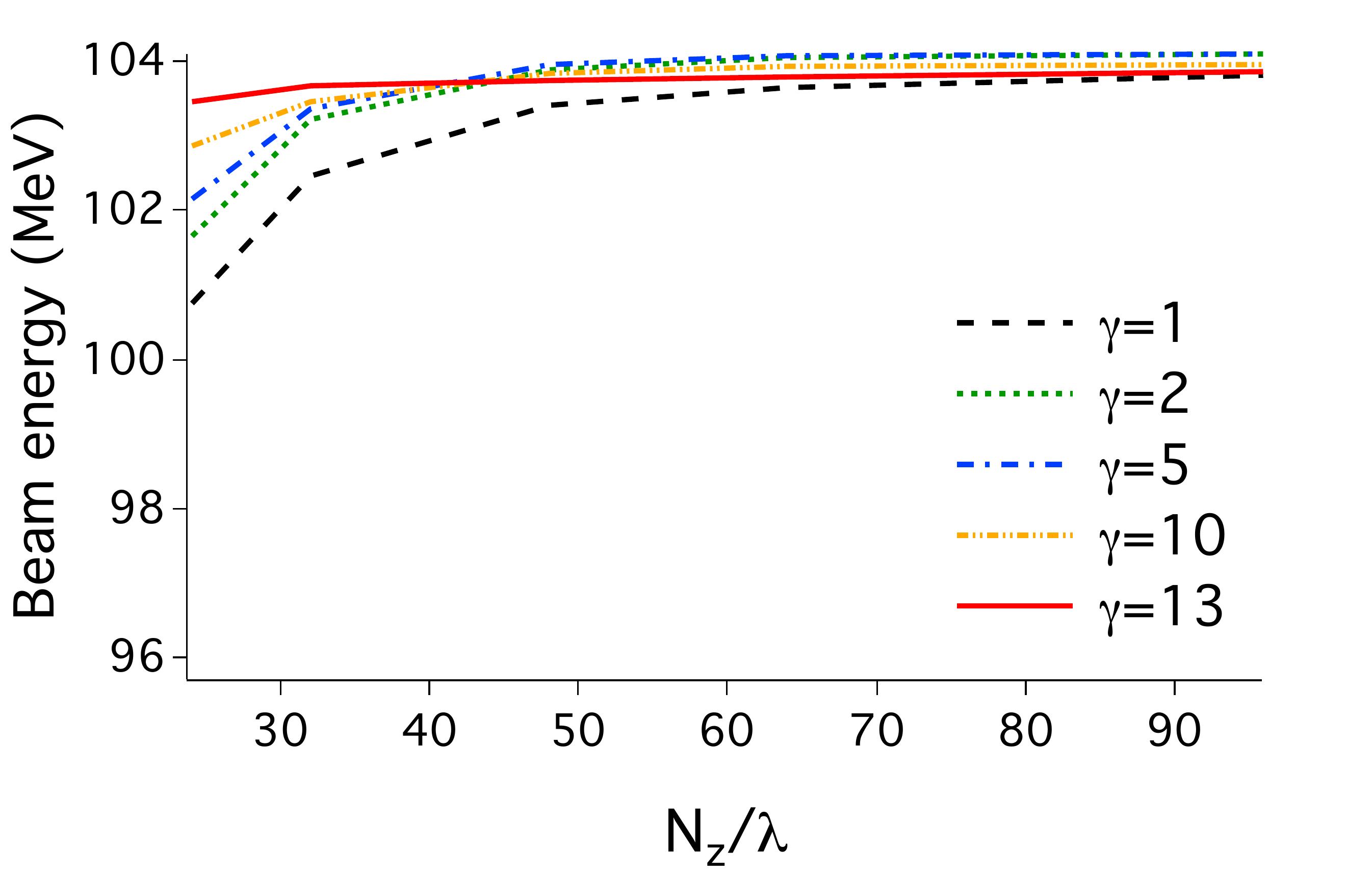}
\includegraphics[trim={0.cm 0.cm 0cm 0.cm},clip,scale=0.25]{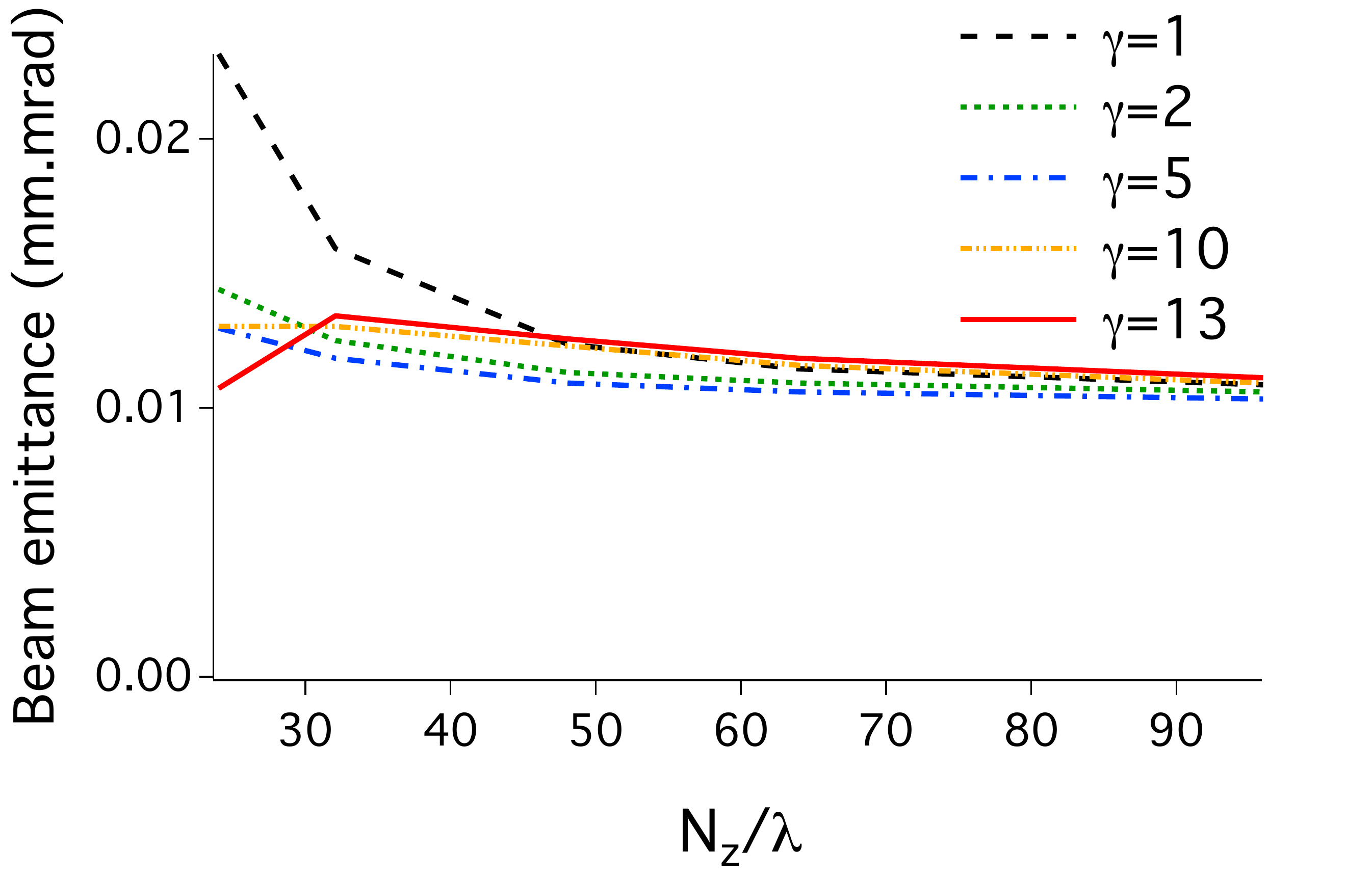}
\caption{\label{fig:beamhist}
Convergence studies on the final energy and emittance of the accelerated 
electron beam at the exit of a laser-driven plasma accelerator, varying the 
longitudinal resolution given by the number of mesh cells $N_z$ per 
laser wavelength $\lambda$. The study was repeated for 
various values of the relativistic factor $\gamma=\{1,2,5,10,13\}$ of 
the Lorentz boosted frame of the simulation.
}
\end{center}
\end{figure}
Given that all PIC simulations are approximations of the real system, it is essential 
to conduct convergence studies with regard to mesh size in every direction, 
number of macroparticles and time step size. 
An example of convergence study, using the final accelerated electron beam 
energy and transverse emittance as metrics, is shown in Fig. \ref{fig:beamhist} 
as the longitudinal resolution is varied between 24 and 96 mesh points 
per laser wavelength. The transverse resolution was kept constant, as well as 
the total number of macroparticles in the accelerated electron beam. The number 
of macroparticles of plasma per cell was also kept constant, hence the total number of plasma 
macroparticles was rising with the longitudinal resolution. The time step was adjusted 
automatically relative to the CFL limit and was thus decreased as the longitudinal resolution 
was decreased.
The study was repeated for various values of the 
relativistic factor $\gamma=\{1,2,5,10,13\}$ of the Lorentz boosted frame of the simulation.

Both the beam energy and emittance converge nicely as the longitudinal resolution 
is increased. However, they do not converge to the same value in each boosted frame. 
This is an indication that the simulation is not converge along all its components, hence 
most likely with regard to the transverse resolution as it was kept constant in this study.
Increasing the transverse resolution did indeed improve the convergence in this case.
Convergence of plasma acceleration simulations can be even more demanding for 
simulations that involve trapping, such as for the study of injection scheme. A 
detailed study of convergence of boosted frame simulations of a laser-driven 
injection scheme is given in \cite{Lee2019}.

\subsubsection{Mesh refinement}
With the standard PIC method, the entire simulation domain is covered with 
a grid at the same resolution everywhere, regardless of the underlying 
physical spatial scales. Hence, the grid cell must be a fraction of the smallest 
spatial feature to be resolved. If the concerned volume is only a 
small subset of the entire domain, it can be very expensive.

There are typically two approaches to enable variable resolution: mesh 
refinement of Cartesian meshes or use of irregular meshes (often 
with the finite element method). The PIC extension to irregular meshes is 
quite complicated, and the tracking of macroparticles through an irregular mesh 
tends to be very expensive. We will thus concentrate here on mesh refinement.

Mesh refinement (MR) is quite common in fluid dynamics but much less in 
electromagnetic simulations, especially when coupled to particles as 
with the electromagnetic PIC method. One reason is that standard MR 
methods, which are based on interpolation of field quantities between grids 
at different resolution, do cause reflection with amplification of electromagnetic modes 
that are resolved on the finer grid but not on the coarser one. To make 
matters worse, the reflection is typically associated with amplification \cite{Vay2001}.
Hence, using standard MR method that are based on interpolation between grids with 
the electromagnetic PIC code leads to unstable methods or need 
a very high level of numerical damping to stabilize the system, with an 
associated cost on physical accuracy. 

Several methods have been proposed to add MR to electromagnetic PIC
without interpolation between fields at different levels of refinement 
\cite{Vaycpc04, VayCSD12, Lo2016, Lo2019}. In \cite{Vaycpc04, VayCSD12}, 
the linearity of Maxwell's equation is used to decouple the field solution 
in the MR area in three parts: parent grid, coarse patch and fine patch. 
In this scheme, the update of Maxwell's equations on the parent grid is 
done without modification of the standard PIC. The effect of MR 
is accounted for by adding a ``correction'' from the combination of the 
field computed on the coarse and fine patches, each surrounded by 
Perfectly Matched Layers \cite{Berengerjcp94} to absorb all outgoing 
waves, avoiding spurious reflections. 

\begin{figure}[ht]
\begin{center}
\includegraphics[trim={0.cm 0.cm 0cm 0.cm},clip,scale=1.]{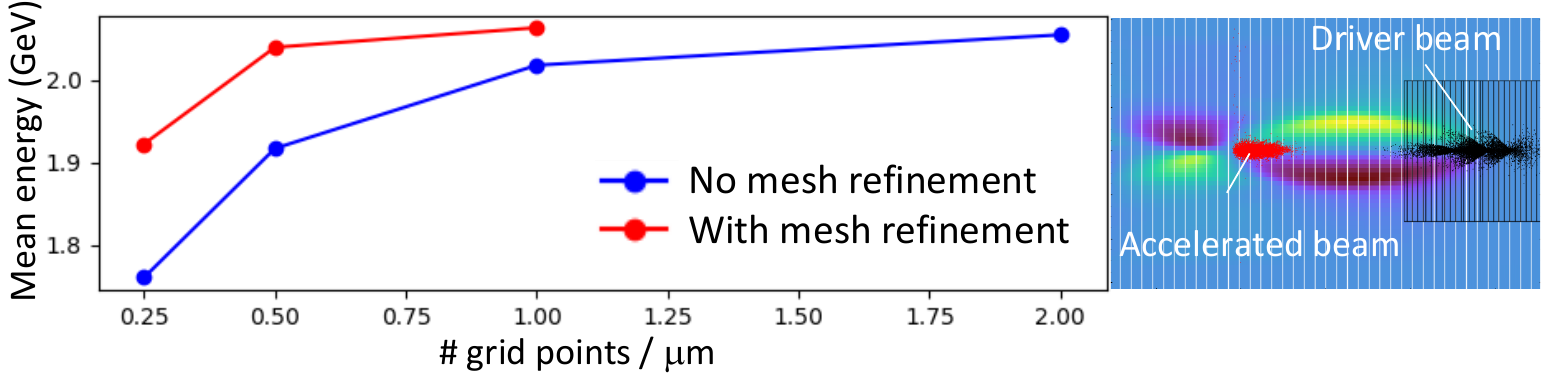}
\caption{\label{fig:PWFAMR}
(left) Final mean energy versus grid resolution of a witness beam at the 
exit of a particle beam-driven plasma accelerator, with (red) or 
without (blue) mesh refinement. (right) Snapshot showing the driver 
beam (black) propagating from left to right, the wake (transverse 
electric field in dark blue and yellow colormap) and the accelerated 
witness electron beam (red). A refinement patch (black grid) is applied 
on the volume occupied by the driver beam.}
\end{center}
\end{figure}
An example of 3-D simulation using this method is shown in Fig. \ref{fig:PWFAMR}, 
where the formation of the wake by a charged particle beam is highly 
sensitive to the grid resolution. A refinement patch is thus applied 
on the volume occupied by the driver beam. The benefit of mesh refinement 
is measured on the dependency of the accelerated beam moments with 
regard to grid resolution on the parent. In this particular example, the 
application of the MR patch around the driver beam only enables 
convergence to a precision nearly identical to the one obtained using 
a fine grid covering the entire simulation domain.

\subsection{High-performance computing \label{Sec:HPC}}
Even using the Lorentz boosted frame technique, mesh refinement, reduced dimensions 
by expanding the fields into a truncated series of azimuthal modes, 
the quasistatic approximation, ponderomotive guiding center (PGC) models, 
or combination of these or other advanced algorithms, the modeling of plasma 
accelerators is computationally very demanding and often calls for 
high-performance computing. This is especially true for the potential future 
application of plasma accelerators to high-energy physics colliders, where 
chains of tens to thousands of multi-GeV plasma accelerators are needed 
to reach multi-TeV energy range in the center of mass of the collision.

The codes that are used for the modeling of plasma accelerators are 
thus in general parallelized, to enable the use of many computer nodes at once. The 
parallelization is usually based on domain decomposition, where the 
simulation domain is decomposed in subdomains, each handled by 
one computer nodes. The decomposition is typically uniform 
Cartesian and the same for fields and particles. At each time step, 
field information is needed from neighboring nodes to compute the finite 
spatial derivatives of fields on the grid, in order to update the Maxwell 
field solver to the next time step. Hence, messages are sent and received between 
neighboring nodes that contain the field information that is needed for the 
update. Similarly, macroparticles may move from one subdomain to 
another and need to be passed via message passing.
Since the distribution of macroparticles may evolve during the simulations 
and may be highly non-uniform (e.g. the plasma electrons in the bubble regime), 
a uniform Cartesian domain decomposition is not always optimal. Hence, 
irregular domain decomposition (which may also be different for fields and 
particles) is needed, with periodic remapping to enable dynamic load balancing 
among computer nodes.

In addition, each computer node may contain multiple processor units, 
including graphical processing units (GPUs). This requires additional 
levels of parallelization and writing of portions of codes using 
annotations and language extensions. The writing of codes that are 
fully optimized with multiple levels of parallelisms and can 
scale to many nodes has thus become very specialized, and 
it is sometimes handled by teams of two or more.

\section{Outlook}

The development of plasma-based accelerators, and in particular high-energy physics colliders, depends critically on high-performance, high-fidelity modeling to capture the full complexity of acceleration processes that develop over a large range of space and time scales. The field will continue to be a driver for pushing the state-of-the-art in the detailed modeling of relativistic plasmas. The modeling of tens to thousands of multi-GeV stages, as envisioned for plasma-based high-energy physics colliders, will require further advances in algorithmic, and depends critically on codes' readiness for the upcoming era of exascale supercomputing. 
This requires coordination within the community with team efforts. 

The emergence of standards for the output data 
(e.g. the open Particle Mesh Data - or \href{https://www.openpmd.org}{openPMD} - standard) and code inputs 
(e.g. the Particle-In-Cell Modeling Interface - or \href{https://github.com/picmi-standard/picmi}{PICMI} - standard), 
if successfully adopted, can pave the way for an integrated ecosystem of new breeds of codes 
that would great simplify the work of many developers and users, thereby leading to speed up of discovery and 
novel designs of plasma accelerators that would not be possible otherwise. 
Standardization of output data will also greatly facilitate a systematic and coordinated development and adoption of ultrafast Machine Learning surrogate models for real-time optimization and control of particle accelerators from simulation (and experimental) data, with a potential for dramatic acceleration of discovery. 

Further down the road, the emerging and largely unexplored 
area of quantum computing offers new opportunities for the modeling of plasma accelerators that are yet to be imagined.

\section*{Acknowledgements}

I would like to thank all my mentors, collaborators, colleagues and students who have 
taught me, pushed me to learn deeper and are continuously 
driving us toward more understanding, novel solutions and new questions.

This work was supported by US-DOE Contract DE-AC02-05CH11231 and by 
the Exascale Computing Project (17-SC-20-SC), a collaborative effort of 
the U.S. Department of Energy Office of Science and the National Nuclear Security Administration.

\appendix

\section{The electromagnetic quasi-static method \label{Sec:QS}}

The electromagnetic quasi-static method was developed earlier than the 
Lorentz boosted frame method, as a way to tackle the large separation of 
length and time scales between the plasma and the driver. 
The quasi-static approximation \cite{SpranglePRL90} takes advantage of the 
facts that (a) the laser or particle beam driver is moving close to the speed 
of light, and is hence very rigid with a slow time response, and (b) the plasma 
response is extremely fast, in comparison to the driver's. The separation 
of the driver and plasma time responses enables a separation in the 
treatment of the two components as follows. 

\begin{figure*}
\begin{centering}
\includegraphics[trim={0.cm 0.cm 0.cm 0.cm},clip,scale=0.6]{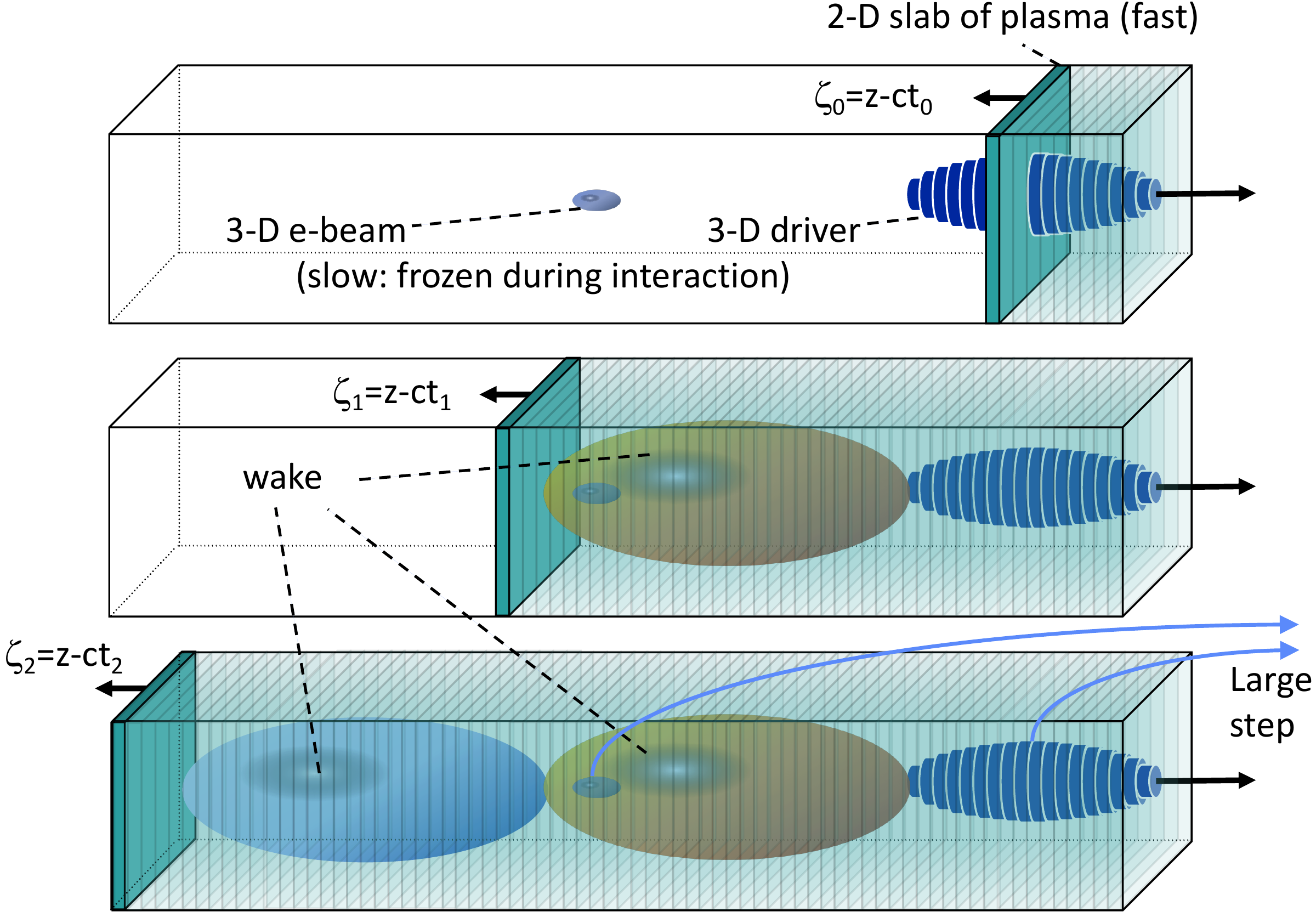}
\par\end{centering}
\caption{\label{fig:Quasistatic} Schematic of the quasi-static method. The 
driver beam and the accelerated electron beam are assumed to be static 
during the time that it takes them to traverse
slices of the plasma column that they are propagating through. This enables a
separation of scales and the evaluation of the response of the plasma to the 
driver, forming the wake, and to the accelerated beam, by following the evolution of a 
2-D slab of the plasma in the coordinates $\zeta=z-ct$.}
\end{figure*}

Assuming the driver at 
a given time and position, its high rigidity enables the approximation 
that it is quasi-static during the time that it takes for traversing a transverse slice 
of the plasma (assumed to be unperturbed by the driver ahead of it). 
The response of the plasma can thus be computed by following the 
evolution of the plasma slice as the driver propagates through it (See Fig. \ref{fig:Quasistatic}). The 
reconstruction of the longitudinal and transverse structure of the wake 
from the succession of transverse slices gives the full electric and 
magnetic field map for evolving the beam momenta and positions 
on a time scale that is commensurate with its rigidity.

Most formulations use the speed-of-light frame, defined as $\zeta=z-ct$, 
to follow the evolution of the plasma slices. Assuming a slice initialized 
ahead of the driver, the evolution of the plasma particles inside the slice 
is given by:

\begin{eqnarray}
\frac{d\mathbf{x}_p}{d\zeta} & = & \frac{d\mathbf{x}_p}{dt}\frac{dt}{d\zeta}=\frac{\mathbf{v}_p}{v_{pz}-c},\\
\frac{d\mathbf{p}_p}{d\zeta} & = & \frac{q}{v_{pz}-c}\left(\mathbf{E}+\mathbf{v}_p\times\mathbf{B} \right).
\end{eqnarray}

The plasma charge and current densities are computed by accumulating 
the contributions of each plasma macro-particle $i$, corrected by the 
time taken by the particle to cross an interval of $\zeta$:

\begin{eqnarray}
\rho_p&=&\frac{1}{\delta x \delta y \delta \zeta}\sum_i \frac{q_i}{1-v_{iz}/c}, \\
\mathbf{J}_p&=&\frac{1}{\delta x \delta y \delta \zeta}\sum_i \frac{q_i \mathbf{v_i}}{1-v_{iz}/c}.
\end{eqnarray}

In contrast, the evolution of a charged particle driver or witness beam (assumed 
to propagate near the speed of light), is given using the standard equations 
of motion:

\begin{eqnarray}
\frac{d\mathbf{x}_{d/w}}{dt} & = &\mathbf{v}_{d/w},\\
\frac{d\mathbf{p}_{d/w}}{dt} & = & q_{d/w}\left(\mathbf{E}+\mathbf{v}_{d/w}\times\mathbf{B} \right),
\end{eqnarray}
while their contributions to the charge and current densities are
\begin{eqnarray}
\rho_{d/w}&=&\frac{1}{\delta x \delta y \delta z}\sum_i q_i, \\
\mathbf{J}_{d/w}&=&\frac{1}{\delta x \delta y \delta z}\sum_i q_i \mathbf{v_i}.
\end{eqnarray}

The electric and magnetic fields are obtained by either solving the equations of the 
scalar and vector potentials in the Coulomb or Lorentz gauge \cite{Morapop1997,Quickpic} 
or directly the Maxwell's equations \cite{LCode,LotovPRSTAB2003,Hipace,Quickpic2} which, under the 
quasi-static assumption 

\begin{equation}
\partial/\partial \zeta = \partial/\partial z = -\partial/\partial ct 
\end{equation}

become 

\begin{eqnarray}
&&\nabla_\bot \times \mathbf{E}_\bot = \frac{\partial B_z}{\partial\zeta}\hat{\mathbf{z}}, \\
&&\nabla_\bot \times E_z\hat{\mathbf{z}} = \frac{\partial \left(\mathbf{B}_\bot-\hat{\mathbf{z}}\times \mathbf{E}_\bot \right)}{\partial\zeta},\\ 
&&\nabla_\bot \times \mathbf{B}_\bot -J_z \hat{\mathbf{z}} = -\frac{\partial E_z}{\partial\zeta}\hat{\mathbf{z}}, \\
&&\nabla_\bot \times B_z\hat{\mathbf{z}} -\mathbf{J}_\bot = -\frac{\partial \left(\mathbf{E}_\bot+\hat{\mathbf{z}}\times \mathbf{B}_\bot \right)}{\partial\zeta}, \\
&&\nabla_\bot \cdot \mathbf{E}_\bot - \rho = -\frac{\partial E_z}{\partial\zeta}, \\
&&\nabla_\bot \cdot \mathbf{B}_\bot = -\frac{\partial B_z}{\partial\zeta}. 
\end{eqnarray}

The set of equations on the potentials or the fields can then be rearranged in a set of
2-D Poisson-like equations that are solved iteratively with the particle motion 
equations. Unlike the Particle-In-Cell method, there is no single way 
of marching the set of equations together and the reader should refer 
to the descriptions of implementations in the various codes for more 
specific details \cite{Morapop1997,Quickpic,LCode,LotovPRSTAB2003,Hipace,Quickpic2}.

\section{The Ponderomotive Guiding Center approximation \label{Sec:PGC}}
For laser pulses with envelopes that are long compared to the laser oscillations, 
it is advantageous to average over the fast laser oscillations and solve the 
laser evolution with an envelope equation \cite{Morapop1997,Turbowave,Quickpic,CowanJCP11,INFERNO}. Assuming a laser pulse in the form 
of an envelope modulating a plane wave traveling at the speed of light,

\begin{eqnarray}
\tilde{A}_\bot = \hat{A}_\bot\left(z,\mathbf{x}_\bot,t\right)\exp{ik_0\zeta}+c.c.,
\end{eqnarray}

the average response of a plasma to the fast laser oscillations can be described 
by a ponderomotive force that inserts into a modified equation of motion:

\begin{eqnarray}
\frac{d\mathbf{p}}{dt} & = & q\left(\mathbf{E}+\mathbf{v}\times\mathbf{B} \right)-\frac{q^2}{\gamma mc^2}\nabla |\hat{A}_\bot|^2,
\end{eqnarray}

with 

\begin{eqnarray}
\gamma = \sqrt{1+\frac{|\mathbf{p}|^2}{m^2c^2}+\frac{2|q\hat{A}_\bot|^2}{m^2c^4}}.
\end{eqnarray}

Most codes \cite{Morapop1997,Turbowave,Quickpic,CowanJCP11} solve the approximate envelope equation

\begin{eqnarray}
\left[\frac{2}{c}\frac{\partial}{\partial t}\left(ik_0+\frac{\partial}{\partial \zeta}\right) + \nabla^2_\bot\right] \hat{A}_\bot \nonumber \\
= \frac{q^2}{mc^2} \Bigg \langle \frac{n}{\gamma}\Bigg \rangle \hat{A}_\bot 
\end{eqnarray}
while the more complete envelope equation 

\begin{eqnarray}
\left[\frac{2}{c}\frac{\partial}{\partial t}\left(ik_0+\frac{\partial}{\partial \zeta}\right) + \nabla^2_\bot - \frac{\partial^2}{\partial t^2}\right] \hat{A}_\bot \nonumber \\
= \frac{q^2}{mc^2} \Bigg \langle \frac{n}{\gamma}\Bigg \rangle \hat{A}_\bot 
\end{eqnarray}
that retains the second time derivative is solved in the code INF\&RNO \cite{INFERNO}. The latter equation is more exact, enabling the accurate simulation of the laser depletion into strongly depleted stages. As noted in \cite{INFERNO}, in order to avoid numerical inaccuracies, or having 
to grid the simulation very finely in the longitudinal direction, it is advantageous to use the polar representation of the laser complex field, namely 
$\hat{A}_\bot(\zeta)=A_\bot(\zeta)\exp[i\theta(\zeta)]$, rather than the more common Cartesian splitting between the real and imaginary parts 
$\hat{A}_\bot(\zeta)=\Re[A_\bot(\zeta)]+\imath\Im[A_\bot(\zeta)]$. As it turns out, the functions $A_\bot(\zeta)$ and $\theta(\zeta)$ are much smoother functions with respect to $\zeta$ than $\Re[A_\bot(\zeta)]$ and $\Im[A_\bot(\zeta)]$, which both exhibit very short wavelength oscillations in $\zeta$, leading to more accurate numerical differentiation along $\zeta$ of the polar representation at a given longitudinal resolution.

\section{Electromagnetic PSATD PIC algorithm in a Galilean frame \label{Sec:Gal}}

In the Galilean coordinates $\vec{x}'$, the equations of particle
motion and the Maxwell equations take the form 
\begin{subequations}
\begin{align}
\frac{d\vec{x}'}{dt} &= \frac{\vec{p}}{\gamma m} - \vgal \label{eq:motion1} \\ 
\frac{d\vec{p}}{dt} &= q \left( \vec{E} +
\frac{\vec{p}}{\gamma m} \times \vec{B} \right) \label{eq:motion2}\\
\left( \Dt{\;} - \vgal\cdot\nab\right)\vec{B} &= -\nab\times\vec{E} \label{eq:maxwell1}\\
\frac{1}{c^2}\left( \Dt{\;} - \vgal\cdot\nab\right)\vec{E} &= \nab\times\vec{B} - \mu_0\vec{J} \label{eq:maxwell2}
\end{align}
\end{subequations}
where $\nab$ denotes a spatial derivative with respect to the
Galilean coordinates $\vec{x}'$. 

Integrating these equations from $t=n\Delta
t$ to $t=(n+1)\Delta t$ results in the following update equations (see
\cite{Lehe2016a} for the details of the derivation):
\begin{subequations}
\begin{align}
\fb^{n+1} &= \theta^2 C \fb^n
 -\frac{\theta^2 S}{ck}i\vec{k}\times \fe^n \nonumber \\
& + \;\frac{\theta \chi_1}{\epsilon_0c^2k^2}\;i\vec{k} \times
                     \fj^{n+1/2} \label{eq:disc-maxwell1}\\
\fe^{n+1} &=  \theta^2 C  \fe^n
 +\frac{\theta^2 S}{k} \,c i\vec{k}\times \fb^n \nonumber \\
& +\frac{i\nu \theta \chi_1 - \theta^2S}{\epsilon_0 ck} \; \fj^{n+1/2}\nonumber \\
& - \frac{1}{\epsilon_0k^2}\left(\; \chi_2\;\mc{\rho}^{n+1} -
  \theta^2\chi_3\;\mc{\rho}^{n} \;\right) i\vec{k} \label{eq:disc-maxwell2}
\end{align}
\end{subequations}
where we used the short-hand notations $\fe^n \equiv
\fe(\vec{k}, n\Delta t)$, $\fb^n \equiv
\fb(\vec{k}, n\Delta t)$ as well as:
\begin{subequations}
\begin{align}
&C = \cos(ck\Delta t) \quad S = \sin(ck\Delta t) \quad k
= |\vec{k}| \label{eq:def-C-S}\\&
\nu = \frac{\vec{k}\cdot\vgal}{ck} \quad \theta =
  e^{i\vec{k}\cdot\vgal\Delta t/2} \quad \theta^* =
  e^{-i\vec{k}\cdot\vgal\Delta t/2} \label{eq:def-nu-theta}\\&
\chi_1 =  \frac{1}{1 -\nu^2} \left( \theta^* -  C \theta + i
  \nu \theta S \right) \label{eq:def-chi1}\\&
\chi_2 = \frac{\chi_1 - \theta(1-C)}{\theta^*-\theta} \quad
\chi_3 = \frac{\chi_1-\theta^*(1-C)}{\theta^*-\theta} \label{eq:def-chi23}
\end{align}
\end{subequations}

Note that, in the limit $\vgal=\vec{0}$,
(\ref{eq:disc-maxwell1}) and (\ref{eq:disc-maxwell2}) reduce to the standard PSATD
equations \cite{HaberICNSP73}, as expected. 

\end{document}